\documentclass[manuscript,screen]{acmart}
\AtBeginDocument{%
  }
  
\acmJournal{TOSEM}

\usepackage{xspace}
\usepackage{comment}
\usepackage{amsmath,amsfonts}
\usepackage{graphicx}
\usepackage{textcomp}
\usepackage{xcolor}
\usepackage{colortbl}
\usepackage{subcaption}
\usepackage[most]{tcolorbox}
\usepackage{booktabs,tabularx,multirow}
\usepackage{pythonhighlight}
\usepackage{hyperref}
\usepackage{lipsum} 
\usepackage{microtype} 
\usepackage{array}
\usepackage{amsthm}
\usepackage{listings}
\usepackage{color}
\usepackage{pgfplots}
\pgfplotsset{compat=1.18}
\usepackage[vlined,ruled,linesnumbered]{algorithm2e}
\usepackage{algpseudocode}
\usepackage{soul}
\usepackage{enumitem}
\usepackage{verbatim}
\usepackage{etoolbox} 
\usepackage{varwidth}

\definecolor{dkgreen}{rgb}{0,0.6,0}
\definecolor{gray}{rgb}{0.5,0.5,0.5}
\definecolor{mauve}{rgb}{0.58,0,0.82}

\theoremstyle{definition} 
\newtheorem{definition}{Definition} 
\def\algosize{\small}

\newtcolorbox[auto counter]{finding}{enhanced,
  attach boxed title to top text left={yshift=-2mm},
  fonttitle=\bfseries, title=Answer to RQ\thetcbcounter}
\newcommand{\Finding}[1]{\begin{finding}#1\end{finding}}

\SetKw{Continue}{continue}

\newcommand{\ourtool}{\textsc{SPARK}\xspace}
\newcommand{\baseline}{baseline \textsc{TCFL}\xspace}

\begin{document}
\title{Similar Pattern Annotation via Retrieval Knowledge for LLM-Based Test Code Fault Localization}

\author{Golnaz Gharachorlu}
\authornote{G. Gharachorlu and M. Panahandeh contributed equally to this work.}
\email{ggharach@uottawa.ca}
\orcid{0000-0002-9891-2811}
\affiliation{%
  \institution{University of Ottawa}
  \country{Canada}
}
\author{Mahsa Panahandeh}
\authornotemark[1]
\email{mpanahan@uottawa.ca}
\orcid{0000-0002-6369-8982}
\affiliation{%
  \institution{University of Ottawa}
  \country{Canada}
}
\author{Lionel C. Briand}
\email{lbriand@uottawa.ca}
\orcid{0000-0002-1393-1010}
\affiliation{%
  \institution{University of Ottawa}
  \country{Canada}
}
\affiliation{ 
\institution{Research Ireland Lero Centre, University of Limerick}
\country{Ireland}}
\author{Ruifeng Gao}
\email{gaoruifeng1@huawei.com}
\affiliation{ 
\institution{Huawei Technologies Co., Ltd.}
\country{China}
}
\author{Ruiyuan Wan}
\email{wanruiyuan@huawei.com}
\orcid{0009-0009-7091-0652}
\affiliation{ 
\institution{Huawei Technologies Co., Ltd.}
\country{China}
}

\begin{abstract}
Software failures remain a major challenge in modern software development, and identifying the code elements responsible for failures is a time-consuming debugging task. While extensive research has focused on fault localization in the system under test (SUT), failures can also originate from faulty system test scripts. This problem, known as Test Code Fault Localization (TCFL), has received significantly less attention despite its importance in continuous integration (CI) environments where large test suites are executed frequently. TCFL is particularly challenging because it typically operates under black-box conditions, relies on limited diagnostic signals such as error messages and partial logs, and involves large system-level test scripts that expand the fault localization search space.
In this paper, we propose \ourtool, a framework that integrates accumulated debugging knowledge from continuous integration (CI) environments into Large Language Model (LLM)-based TCFL. Given a newly observed failing test case, \ourtool retrieves similar fault-labeled test cases from a debugging knowledge corpus and selectively annotates suspicious lines of the failing test based on their similarity to previously observed fault patterns. These annotations guide the LLM's reasoning while maintaining scalability and avoiding the prompt-length explosion common to naive retrieval-augmented approaches.
We evaluate \ourtool on three industrial datasets containing real-world faulty Python test cases from different software products. The results show that \ourtool consistently improves fault localization effectiveness compared to the existing LLM-based TCFL baseline while maintaining comparable inference cost and token usage. In particular, the approach advances the state of the art by identifying more correct faulty locations in complex test cases containing multiple faults.
\end{abstract}



\maketitle

\section{Introduction}\label{sec:introduction}
Software failures remain a fundamental challenge in modern software development~\cite{10.1145/3779132}. As software systems grow in scale and complexity, diagnosing the root causes of failures becomes increasingly time-consuming and costly~\cite{buglocalization14,buglocalization22,kept25}. Empirical studies indicate that developers often spend a substantial portion of debugging effort identifying the code locations responsible for observed failures, even before any repair can begin~\cite{FL_study17}. For example, a human study by Boehm et al.~\cite{FL_study17} shows that localizing a moderately difficult fault requires more than 30 minutes for a proficient developer, while difficult faults may take up to 55 minutes. Consequently, fault localization (FL), the task of identifying code elements most likely responsible for a failure~\cite{wong2016survey}, plays a central role in improving software reliability and developer productivity~\cite{10.1145/3779132}.

A large body of research~\cite{DBLP:journals/ieicetd/ZhengHCYFX24,DBLP:journals/jss/RaselimoF24,DBLP:journals/access/SarhanB22,de2016spectrum,zakari2020spectrum,DBLP:conf/icse/YangGMH24,DBLP:journals/corr/abs-2403-16362} has focused on FL within the System Under Test (SUT), where failures may manifest during testing but originate from faults in the production code. This setting is commonly referred to as SUT fault localization (SUTFL)~\cite{saboor2025black}. A widely adopted class of SUTFL techniques is spectrum-based fault localization (SBFL)~\cite{DBLP:journals/ieicetd/ZhengHCYFX24,DBLP:journals/jss/RaselimoF24,DBLP:journals/access/SarhanB22,de2016spectrum,zakari2020spectrum}, which collects execution coverage information from test cases and identifies suspicious code elements based on statistical correlations between test outcomes and execution behavior. Beyond SBFL, more recent research has explored machine learning (ML)-based approaches that learn system dependencies from diverse artifacts, such as source code, fault reports, and execution logs, to improve localization accuracy~\cite{traceability2018,NP-CNN,MD-CNN,MRAM,DBLP:conf/sigsoft/LouZDLSHZZ21}. These techniques often employ neural networks to capture the SUT's structural and semantic characteristics. More recently, with the emergence of large language models (LLMs), researchers have begun leveraging these models' reasoning capabilities to analyze code and failure artifacts for improved FL performance~\cite{DBLP:conf/icse/YangGMH24,kept25, 10989036}.

However, failures observed in testing environments are not always caused by faults in the system under test. In many cases, the test code itself is faulty~\cite{Vahabzadeh2015empirical}. Test scripts may contain incorrect assertions, outdated expectations, fragile assumptions about the system state, or errors introduced during automated test generation. These issues give rise to Test Code Fault Localization (TCFL), a problem setting in which both the failure and its root cause reside within the test code rather than the SUT~\cite{saboor2025black}. Despite its practical importance, particularly in large continuous integration (CI) pipelines where thousands of tests are executed daily, TCFL has received significantly less attention than traditional SUTFL.

TCFL poses several unique challenges that distinguish it from traditional SUTFL. First, TCFL often operates under black-box conditions, where the SUT's source code is unavailable to the test analysis pipeline. This is because system-level testing typically evaluates software through external interfaces and observable behavior rather than internal implementation details~\cite{10.5555/2161638}, thereby complicating fault reasoning~\cite{saboor2025black}. Second, test failures typically provide limited diagnostic signals, such as error messages and partial execution logs. Unlike traditional FL settings, where detailed execution traces and internal program states may be available, failures observed during testing often do not expose a complete trace of the SUT's execution or its internal runtime state. Moreover, there are no test cases for the test code, making execution-based FL techniques such as SBFL inapplicable in TCFL. Consequently, the available information is often insufficient to accurately determine the underlying root cause of the failure~\cite{DBLP:journals/software/BaudryFJT05,DBLP:journals/access/GuptaSP19}. Third, modern test suites often include large, complex system-level test scripts that interact with multiple components of the SUT. These characteristics substantially expand the search space that must be explored during FL. As a result, these characteristics severely limit the effectiveness of conventional FL techniques when applied directly to test code.

Saboor et al.~\cite{saboor2025black} present, to the best of our knowledge, the only existing approach specifically designed for TCFL. Their work explores the use of LLMs for TCFL, leveraging their ability to reason over test code and natural-language error messages. The proposed method operates in a black-box setting with respect to the SUT and enriches the input context by generating estimated execution traces of the test code from the available failed execution logs. While promising, the approach largely relies on vanilla prompting, where the LLM receives only the failing test code and its associated error message as input, with additional efficiency improvements achieved through trace estimation. Such limited contextual information may lead to suboptimal localization performance, particularly for complex test cases in which the root cause cannot be inferred directly from the failure message. This limitation becomes especially evident in scenarios involving multiple faulty locations, where pruning candidate lines based solely on estimated traces may exclude relevant fault locations. More importantly, this approach does not leverage a valuable source of domain-specific debugging knowledge that naturally accumulates in industrial development environments through the diagnosis and resolution of test failures.

In practice, continuous integration (CI) pipelines repeatedly execute builds and tests, generating failures that developers investigate and resolve throughout the lifetime of a system~\cite{duvall2007continuous, saff2003reducing}. Each diagnosed failure caused by faults in the test scripts contributes to building valuable knowledge about recurring faulty patterns in test code. Over time, these resolved failures form a growing corpus of fault-labeled debugging examples, which can provide strong contextual signals for diagnosing new failures. 

However, effectively incorporating such debugging knowledge into LLM-based FL remains an open challenge. Beyond identifying and selecting relevant contextual information from this corpus, naively injecting entire debugging examples into prompts leads to excessive token consumption and poor scalability. Additionally, fine-tuning LLMs on project-specific debugging datasets is often impractical due to limited labeled data, high computational cost, and maintenance overhead~\cite{DBLP:conf/icse/NashidSM23}. Moreover, models fine-tuned on project-specific data often struggle to generalize effectively to new projects.

In this paper, we propose \ourtool, 
$\mathcal{S}$imilar $\mathcal{P}$attern $\mathcal{A}$nnotation via $\mathcal{R}$etrieval $\mathcal{K}$nowledge, an end-to-end framework for TCFL that systematically integrates accumulated debugging knowledge into the FL reasoning process of LLMs. 
Rather than embedding raw debugging knowledge directly into the prompt, \ourtool adapts a lightweight annotation-based retrieval mechanism. Given a failing test case, referred to as the \emph{query test case}, \ourtool retrieves only one or a few similar fault-labeled test cases from the debugging knowledge, extracts their faulty lines, and selectively annotates lines in the query test case that are textually similar to these previously observed faulty patterns. These annotations serve as structured guidance signals that help the LLM focus its reasoning on a smaller and more relevant subset of candidate code elements. This design offers two key advantages. First, it incorporates domain-specific debugging knowledge without requiring model retraining, enabling practical deployment with off-the-shelf LLMs. Second, it preserves scalability by avoiding the large prompt sizes associated with naive retrieval-augmented approaches and alternative strategies such as in-context learning (ICL)~\cite{dong-etal-2024-survey,ICL22} or few-shot learning~\cite{DBLP:journals/csur/WangYKN20} when the examples are large. As a result, \ourtool enables more accurate and efficient FL in realistic CI environments where large test suites and frequent failures are common.

To systematically investigate key design choices for integrating debugging knowledge into LLM-based TCFL, we further (1) examine practical data availability scenarios for building the debugging knowledge corpus, (2) show the importance of extracting relevant contextual information, (3) demonstrate the necessity of a lightweight mechanism for augmenting prompts with the extracted knowledge, and (4) explore the impact of different prompt templates~\cite{errica2024did,wang2024prompt,hua2025flaw} on FL effectiveness.

We evaluate \ourtool on three industrial datasets containing varying numbers of real-world faulty Python test cases, each corresponding to the testing and verification of a different software product (SUT). The results demonstrate that our approach significantly improves FL effectiveness compared to the existing LLM-based TCFL baseline~\cite{saboor2025black}, with minimal computational overhead. At the line level, \ourtool improves top-1 Precision and Hit by 10.1--19.1 percentage points (pp) and Recall by 6.8--11.9 pp compared to the baseline. These gains are consistent across different experimental setups and datasets. For instance, at the line level, Hit@3 increases by 17.7--25.5 pp across the three datasets, achieving a maximum value of 85.7\% on the best-performing dataset and 57.4\% on the lowest-performing dataset. At coarser granularity levels, Hit@3 improves by 9.2--25 pp, reaching up to 100\%. These improvements translate into a higher number of correctly identified faulty test code elements, particularly in test cases containing multiple fault locations. At the same time, the approach remains scalable in terms of prompt token usage compared to the baseline, requiring only 10--60 additional input tokens per test case, with inference times nearly identical to those of the baseline. Additional experiments further demonstrate the robustness of the \ourtool design under different configuration settings and experimental scenarios.

The contributions of this paper are summarized as follows:

\begin{enumerate}
\item \textbf{Problem formulation for TCFL in CI environments.}
We characterize the practical setting of \emph{Test Code Fault Localization (TCFL)} in continuous integration (CI) environments and highlight the opportunity to leverage accumulated debugging knowledge from test failure diagnosis. We show that effectively leveraging this knowledge for TCFL remains a challenging, largely unexplored problem.

\item \textbf{Retrieval-augmented fault localization framework.}
We propose \ourtool, an end-to-end framework for LLM-based TCFL that adapts the principles of retrieval-augmented generation (RAG). To the best of our knowledge, \ourtool is the first approach to systematically leverage accumulated debugging knowledge to guide LLM reasoning for TCFL, while addressing key practical design considerations for deployment in real-world CI environments.

\item \textbf{Annotation-based context integration.}
We introduce a lightweight annotation-based mechanism that selectively highlights suspicious lines in failing test cases by comparing them to previously observed faulty patterns. This design improves the effectiveness of LLM-based TCFL while maintaining scalability by avoiding the prompt-size explosion associated with naive retrieval-based approaches.

\item \textbf{Comprehensive empirical evaluation.}
We evaluate \ourtool on three industrial datasets containing real-world faulty Python test cases from different software products. The results show that \ourtool significantly improves FL effectiveness compared to the existing LLM-based TCFL baseline, while maintaining comparable efficiency.

\item \textbf{Tool and data availability.}
We make our tool implementation and the processed dataset (in the form of embeddings) available online\footnote{\underline{
Available after review completion.}}. The embeddings are provided instead of the raw data due to privacy constraints imposed by our industrial partner. They preserve the essential characteristics of our datasets, including the accumulated debugging knowledge. Together with the implementation, the provided resources support transparency, facilitate comparative evaluations within the same setting, and enable future extensions of the proposed approach.
\end{enumerate}

The remainder of this paper is organized as follows. Section~\ref{sec:background} reviews the necessary background and formalizes the TCFL problem. It also presents a motivating experiment illustrating why naive RAG-based TCFL approaches are inadequate in practical CI environments. Section~\ref{sec:method} formalizes the proposed \ourtool framework. Section~\ref{sec:evaluation} describes the experimental design and presents the evaluation results. Section~\ref{sec:discussion} analyzes the impact of different design choices in our approach. Section~\ref{sec:threats} discusses threats to validity and the measures taken to mitigate them. Related work is presented in Section~\ref{sec:rw}. Finally, Section~\ref{sec:conclusion} concludes the paper and outlines directions for future work.

\section{Background and Motivation}\label{sec:background}
In this section, we introduce the fundamental concepts, terminology, and background necessary to understand our approach and its underlying motivation.

\subsection{Test Code Fault Localization}\label{llm::tcfl}
In modern continuous integration (CI) pipelines, test failures occur frequently and must be diagnosed quickly to maintain developer productivity and support timely regression detection~\cite{duvall2007continuous, saff2003reducing}. Such failures often arise when changes to the SUT break existing tests, causing them to fail against the updated system~\cite{yaraghi2025automated}. Faulty tests may also originate from automated test generation, where inferred assertions or overfitting to current program behavior result in fragile or incorrect tests~\cite{schafer2023empirical, yang2024empirical}. Test Code Fault Localization (TCFL) focuses on identifying faulty elements, such as files, functions, or lines, within the \emph{test code} itself that cause test failures, rather than faults in the system under test (SUT)~\cite{saboor2025black, Vahabzadeh2015empirical}. 

TCFL is typically black-box with respect to the SUT~\cite{saboor2025black}. Although execution logs may reveal partial execution signals, such as references to executed SUT components or line numbers, the SUT source code, its dependencies, and internal semantics are often unavailable to the test analysis pipeline. This limitation arises because TCFL focuses on diagnosing faults in the test code rather than in the SUT itself. System-level testing is usually conducted through the SUT's public interfaces and observable outputs, without relying on implementation details~\cite{10.5555/2161638}. Moreover, in many practical scenarios, the SUT may be a deployed or proprietary system, or its source code may include sensitive information, making direct access difficult or undesirable~\cite{katyal2018paradox, raemaekers2011exploring, DBLP:conf/icsm/Wang0HSX0WL20}. Consequently, the test analysis pipeline must reason about test failures solely from execution outcomes and observable signals, which lack the semantic context required for white-box reasoning about SUT behavior, thereby making FL more challenging. Moreover, test failures typically provide limited diagnostic information~\cite{DBLP:journals/software/BaudryFJT05,DBLP:journals/access/GuptaSP19}, often restricted to error messages and partial logs, which further complicates root-cause analysis. Finally, modern test suites often include large, complex system-level test scripts that interact with multiple SUT components~\cite{10254752,PONCE2025107870}, substantially expanding the search space to be explored during FL. These challenges make TCFL fundamentally different from, and more challenging than, traditional SUTFL, which commonly relies on fine-grained execution traces, coverage data, and white-box program analysis~\cite{nidhra2012black}. 

\begin{figure}[t]
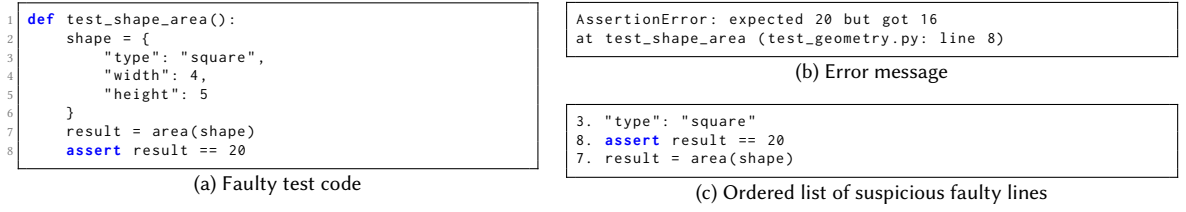

   \centering
   \noindent
   \begin{subfigure}{0.44\linewidth}
   \begin{lstlisting}[language=Python,
       basicstyle=\ttfamily\scriptsize\color{black},
       keywordstyle=\color{blue}\bfseries,
       commentstyle=\color{black},
       stringstyle=\color{black},
       showstringspaces=false,
       numbers=left,
       numberstyle=\tiny\color{gray},
       stepnumber=1,
       numbersep=5pt,
       frame=single,
       breaklines=true,
       breakatwhitespace=true,
       tabsize=4,
       captionpos=b
   ]
def test_shape_area():
    shape = {
        "type": "square",
        "width": 4,
        "height": 5
    }
    result = area(shape)
    assert result == 20
   \end{lstlisting}
   \vspace{-1em}
   \caption{Faulty test code}
    \vspace{-1.6em}
   \end{subfigure}%
   \hfill
   \begin{subfigure}{0.52\linewidth}
       \centering
       \begin{subfigure}{\linewidth}
       \begin{lstlisting}[language=Python,
           basicstyle=\ttfamily\scriptsize\color{black},
           keywordstyle=\color{blue}\bfseries,
           commentstyle=\color{black},
           stringstyle=\color{black},
           showstringspaces=false,
           numbers=none,
           numberstyle=\tiny\color{gray},
           stepnumber=1,
           numbersep=5pt,
           frame=single,
           breaklines=true,
           breakatwhitespace=true,
           tabsize=4,
           captionpos=b
       ]
AssertionError: expected 20 but got 16
at test_shape_area (test_geometry.py: line 8)
       \end{lstlisting}
          \vspace{-1em}
       \caption{Error message}
       \end{subfigure}

       \vspace{0.5em}

       \begin{subfigure}[c]{\linewidth}
       \begin{lstlisting}[language=Python,
           basicstyle=\ttfamily\scriptsize\color{black},
           keywordstyle=\color{blue}\bfseries,
           commentstyle=\color{black},
           stringstyle=\color{black},
           showstringspaces=false,
           numbers=none,
           numberstyle=\tiny\color{gray},
           stepnumber=1,
           numbersep=5pt,
           frame=single,
           breaklines=true,
           breakatwhitespace=true,
           tabsize=4,
           captionpos=b
       ]
3. "type": "square"
8. assert result == 20
7. result = area(shape)
       \end{lstlisting}
          \vspace{-1em}
       \caption{Ordered list of suspicious faulty lines}
       \end{subfigure}
   \end{subfigure}

   \caption{A synthetic example showing a faulty test case, its error message, and the list of lines ranked by their likelihood of being faulty. Note that the reported error location (line 8) does not necessarily coincide with the true fault location (line 3).}
   \label{fig::simple::running}
\end{figure}

Consequently, TCFL generally relies only on the test source code and its associated runtime failure information, such as error messages and logs, and may need to reason under an incomplete or estimated execution context. Given an observed failure during the execution of a faulty test case, hereafter referred to as the \emph{query test case}, the goal of TCFL is to identify the root cause(s) of the test failure (faults) within the query test case. FL techniques commonly adopt ranking-based strategies~\cite{DBLP:conf/icse/JonesHS02, DBLP:conf/prdc/AbreuZG06, 2014ITR}, producing a list of code locations ordered by their likelihood of being faulty. These locations, referred to as \emph{elements}, can be defined at varying levels of code \emph{granularity}, including files, functions, basic blocks in control flow graphs (CFGs)~\cite{cfgbook}, and individual lines of code. \autoref{fig::simple::running} illustrates a synthetic example of a faulty test case with an assertion error on line 8, along with its ranked list of suspicious elements at line-level granularity. In this example, the failure observed at line 8 of the test code is caused by an incorrect test input defined at line 3, where the shape is specified as "square" despite providing rectangular dimensions. A desirable TCFL approach should rank the true fault location (line 3) at the top of the suspiciousness ranking, while assigning lower ranks to less likely locations.

\subsection{LLM-Based Test Code Fault Localization}

With the widespread adoption of large language models (LLMs) in software engineering, their application to FL has gained increasing attention~\cite{DBLP:journals/corr/abs-2403-16362, FlexFL, SoapFL, saboor2025black}. Owing to their strong reasoning capabilities and natural language generation, LLMs are well-suited to analyze failure contexts and produce human-readable diagnostic outputs~\cite{codebert2020}. LLM-based TCFL approaches leverage these capabilities to automatically locate faults in a query test case via prompt engineering, in which the LLM is provided with the faulty test code, the associated failure message, and task-specific instructions to guide the FL process.

\begin{figure}[!t]
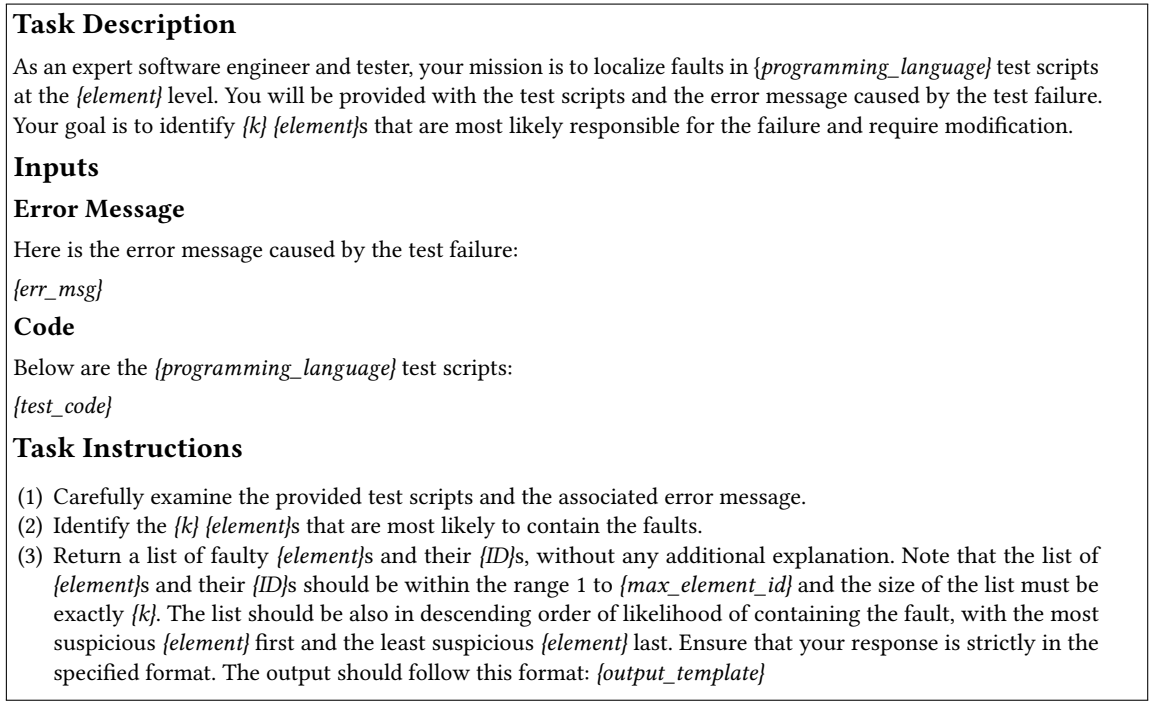

  \centering
  \scriptsize
  \framebox[\columnwidth][l]{\parbox{0.95\columnwidth}{
  \Large \textbf{Task Description} \vspace{0.4em} \\
  \normalsize As an expert software engineer and tester, your mission is to localize faults in \{\emph{programming\_language\}} test scripts at the \emph{\{element\}} level. You will be provided with the test scripts and the error message caused by the test failure. Your goal is to identify \emph{\{k\}} \emph{\{element\}}s that are most likely responsible for the failure and require modification.\vspace{0.6em}\\
 \Large \textbf{Inputs} \vspace{0.4em} \\
\large \textbf{Error Message}
 \vspace{0.4em}\\
\normalsize Here is the error message caused by the test failure:\vspace{0.4em}\\
  \emph{\{err\_msg\}}\vspace{0.4em}\\
\large \textbf{Code}\vspace{0.4em}\\
\normalsize Below are the \emph{\{programming\_language\}} test scripts:\vspace{0.4em}\\
   \emph{\{test\_code\}}\vspace{0.6em}\\
\Large\textbf{Task Instructions}\vspace{0.4em}
\normalsize
\begin{enumerate}[leftmargin=15pt]
\item Carefully examine the provided test scripts and the associated error message.
    \item Identify the \emph{\{k\}} \emph{\{element\}}s that are most likely to contain the faults.
    \item Return a list of faulty \emph{\{element\}}s and their \emph{\{ID\}}s, without any additional explanation. Note that the list of \emph{\{element\}}s and their \emph{\{ID\}}s should be within the range 1 to \emph{\{max\_element\_id\}} and the size of the list must be exactly \emph{\{k\}}. The list should be also in descending order of likelihood of containing the fault, with the most suspicious \emph{\{element\}} first and the least suspicious \emph{\{element\}} last. Ensure that your response is strictly in the specified format. The output should follow this format: \emph{\{output\_template\}}
\end{enumerate}
  }}
  \caption{The \baseline's prompt template for test code fault localization~\cite{saboor2025black}. Text enclosed in curly braces (\{ \}) represents variable placeholders dynamically filled during the prompting process.}
  \label{fig:prmpt_template}
\end{figure}

To the best of our knowledge, the TCFL approach proposed by Saboor et al.~\cite{saboor2025black}, which we refer to as the \emph{\baseline} throughout this paper, is the only existing method explicitly designed to address the TCFL problem. This baseline implements a fully automated LLM-based solution for TCFL.
\autoref{fig:prmpt_template} presents an example prompt template used in this approach~\cite{saboor2025black}. The template employs role-based prompting~\cite{DBLP:conf/naacl/KongZCLQSZWD24}, assigning a specific role to the LLM and providing a task description to guide the FL process. The prompt inputs comprise the error message and the faulty test code, along with instructions directing the LLM to analyze the information and return a ranked list of suspicious code elements, ordered by decreasing likelihood of faultiness.

Formally, the \baseline approach is defined as follows.

\begin{definition}[\emph{\textbf{Baseline TCFL}}]\label{def::tcfl}

Given a query test \(\tau_f\), \baseline leverages a set of instructions \(I\) to produce a list of \( k \) candidate elements ranked in descending order of suspiciousness, as inferred by the LLM through a probability function \( P: E \rightarrow [0, 1] \):

\[
\text{TCFL}_\text{base}(\tau_f, I, k) = \{ e_1, e_2, \dots, e_k \}
\]
\\
where \(\tau_f\) consists of a set of lines \(\{l_1, l_2, \dots, l_n\}\) and an associated error message, \(E\) denotes the set of all candidate elements at the chosen granularity,  \( e_i \in E \) for \( i = 1, 2, \dots, k \), \(k \leq n\), and \( P(e_k) \leq P(e_{k-1}) \leq \dots \leq P(e_2) \leq P(e_1) \). Depending on the selected code granularity, each element \(e_i\) may correspond to a function, a CFG block, or an individual line within \(\tau_f\).
\end{definition}

Relying solely on the test code and its associated error message, the \baseline approach provides the LLM with limited contextual guidance. In particular, \baseline prompts the LLM with either the full test code or a pruned subset of executed lines, implicitly assuming that all candidate elements are equally likely to be faulty. As a result, the LLM relies primarily on a single diagnostic signal, the error message, to identify the root cause in potentially large and complex test cases.

Our experimental results (see \autoref{sec:evaluation}) indicate that this single-modality guidance limits the effectiveness of the \baseline approach across different datasets, especially for system-level test scripts. For example, in many cases, \baseline identifies only the code locations associated with the error message as the top-ranked faulty elements, which is often ineffective, especially in datasets where test cases may contain multiple faulty elements. 

As mentioned, system-level test scripts typically exercise large portions of the SUT and involve complex interactions between test code, APIs, and system components. In these settings, additional contextual information is required to enable the LLM to reason more effectively and to distinguish faulty code from non-faulty code. 
At the same time, it is crucial to carefully select which contextual information to provide and how to integrate it into the LLM, given its context-length constraints. In the absence of a white-box view of the SUT, this observation motivates using \emph{debugging knowledge} as an external source of contextual information that can be selectively retrieved and incorporated to better support the TCFL process.

\subsection{Retrieval-Augmented Generation (RAG) for Fault Localization}

Recent LLM-based system understanding and reasoning methods have increasingly moved beyond vanilla prompts toward augmenting prompts with additional contextual information---such as repository artifacts, change sets, and logs---to improve the effectiveness of LLM reasoning for tasks such as patch generation~\cite{rapgen23}, program repair~\cite{RAGFix}, and SUTFL~\cite{RAG-SUTFL-Du, RAG-SUTFL-Shi}. This trend is largely driven by Retrieval-Augmented Generation (RAG)~\cite{rag-work}, which enhances LLM responses by retrieving relevant external information at inference time and incorporating it into the prompt.

RAG is particularly useful because vanilla LLMs may lack the domain and project-specific knowledge needed to reason reliably about failures across diverse systems and environments. This challenge becomes more pronounced in settings where: (i) production or CI constraints enforce a black-box setting with limited access to SUT internals or repository-wide dependencies; (ii) fine-tuning is impractical due to cost, data privacy, or operational constraints; and (iii) the target system and its failure modes are unlikely to have been observed by the LLM during pretraining. In industrial contexts, this last point is especially important, as evaluations on widely used open-source projects may overestimate real-world performance when the model has potentially seen similar artifacts during pretraining.

In such settings, a key reliable source of context is debugging knowledge accumulated within the CI pipeline. In practice, failures often recur~\cite{Li2021APImisuse, parry2025systemic, Ain2019codeclone, Vahabzadeh2015empirical}, and the process of diagnosing and correcting them is not necessarily time-dependent: some faults are quickly identified, fixed, and labeled, while others may remain unresolved due to factors such as ownership boundaries, severity, engineering cost, or limited expertise~\cite{Hu2014bugtriage}. The set of previously localized and labeled faults, therefore, constitutes a valuable, growing knowledge base that can support the diagnosis of new failures.

Motivated by this observation, we retrieve additional context from external accumulated debugging knowledge and incorporate it into the LLM prompt. Specifically, we (i) retrieve the most similar available faulty labeled test case(s) to the \emph{query test case}, and (ii) annotate the code lines in the query test case that closely resemble the faulty lines in the retrieved test case(s). Since both effectiveness and efficiency are critical, it is crucial to carefully determine which information to retrieve and how to present it to the LLM in a compact, effective manner. We detail our retrieval strategy, context construction, and the components of our approach that handle different practical scenarios in Section~\ref{sec:method}.

\subsection{Limitations of Naive RAG for Test Code Fault Localization: A Motivation Study}
\label{sec:motivation}
Although RAG-based approaches address the lack of contextual information and provide a principled mechanism for incorporating external knowledge into LLM-based reasoning, naively applying RAG to TCFL is often inefficient and impractical in industrial settings. In contrast, selectively retrieving and adopting only highly relevant information can substantially reduce token usage, improve scalability, and potentially enhance localization effectiveness. In this section, we empirically demonstrate the limitations of a naive RAG-based TCFL pipeline under realistic deployment constraints.

Given a query test $\tau_f$, a typical naive RAG-based TCFL approach would retrieve the most relevant labeled faulty test case(s) from the debugging knowledge as additional context. The retrieved test code, error messages, and labeled faulty elements would then be integrated into the LLM's prompt. Compared to the \baseline's prompt template (\autoref{fig:prmpt_template}), this approach also requires additional instructions to guide the LLM in reasoning over the retrieved examples while localizing faults in $\tau_f$. However, this design faces three major practical challenges:

\begin{enumerate}
    \item \textbf{Computational and context constraints.} 
    Processing long inputs increases GPU memory usage, inference latency, and computational overhead due to the transformer’s attention mechanism~\cite{vaswani2017attention}. Embedding entire retrieved test cases significantly increases prompt length, often approaching or exceeding the context limits of commonly used LLMs.

    \item \textbf{Operational cost under frequent invocation.} 
    Continuous integration pipelines run frequently, and invoking an LLM with extended prompts for each failure incurs substantial cumulative token consumption and inference costs, which can become economically impractical at scale.

    \item \textbf{Scalability with respect to test size.} 
    System-level test scripts are often large and complex. A naive RAG strategy that includes fully retrieved test cases causes the prompt size to grow proportionally with both the query test and the retrieved context, leading to poor scalability as test cases increase in size.
\end{enumerate}

\begin{table*}[!t]
\centering
\scriptsize
\resizebox{\textwidth}{!}{
\begin{tabular}{l cccc cccc cccc cccc}
\toprule
\multirow{2}{*}{Dataset} 
& \multicolumn{4}{c}{Query Test Case Tokens (\#)} 
& \multicolumn{4}{c}{Full Retrieved Context Tokens (\#) } 
& \multicolumn{4}{c}{Naive RAG Prompt Tokens (\#) } 
& \multicolumn{4}{c}{\ourtool Prompt Tokens (\#)} \\
\cmidrule(lr){2-5}
\cmidrule(lr){6-9}
\cmidrule(lr){10-13}
\cmidrule(lr){14-17}
& Min & Median & Mean & Max 
& Min & Median & Mean & Max
& Min & Median & Mean & Max
& Min & Median & Mean & Max \\
\midrule

DS1 & 1,215.0  & 5,636.0 & 6,570.0 & 29,929.0 &
       1,371.0 & 5,596.0 & 6,560.4 & 29,992.0 &  
         3,204.0   & 11,877.0 & 13,551.5  & 55,595.0 &  
          1,645.0  & 6,434.0 &  7,495.1 & 32,820.0  \\

DS2 & 1,517.0 & 6,609.5 & 8,218.4 &  20,840.0&
       1,588.0     & 6,642.0 & 8,095.3 & 19,707.0 &  
        3,724.0    & 13,893.5 & 16,734.7 & 36,057.0 &  
           1,986.0 & 7,538.0 & 9,486.9 &24,397.0  \\

DS3 &  1,942.0 & 4,434.0 & 5,568.6 &  16,171.0&
         2,490.0   & 4,063.0 & 4,796.1 & 12,135.0 &  
          4,930.0  & 9,172.0 & 10,785.7 & 24,524.0 &  
           2,411.0 & 5,104.0 & 6,284.9 & 17,667.0 \\

\bottomrule
\end{tabular}
}
\caption{Token usage comparison across three datasets for query test cases, retrieved context, naive RAG-based TCFL prompts, and \ourtool prompts. \ourtool reduces token consumption substantially while maintaining effective context.}
\label{table::benchmark::motiv}
\end{table*}

To quantify this overhead, we measure the token usage of a typical naive RAG-based TCFL approach that retrieves the single most similar test case and embeds its entire context directly into the prompt, across three industrial system-level test script datasets (DS1--DS3), and compare it with token usage of \ourtool, which instead annotates only selected lines in \(\tau_f\) to guide the LLM's attention toward lines that are similar to the retrieved test case. 

~\autoref{table::benchmark::motiv} reports token statistics for the query test (its code and error message), the raw retrieved context, including the most similar test code, its error messages, and labeled faulty elements, as well as the total prompt size under both naive RAG-based TCFL and \ourtool.
The results show that naive RAG substantially increases token usage, often exceeding practical context window limits of open-source LLMs. In contrast, \ourtool selectively retrieves and incorporates only highly relevant lines, reducing prompt token usage by nearly half on average across datasets (44.7\%, 43.3\%, 41.7\%), clearly demonstrating the practical limitations of naive RAG for industrial deployments. In large-scale CI environments, these token savings translate directly into faster LLM inference, lower operational costs, and the ability to handle larger or more complex test scripts. While efficiency is not the primary goal of \ourtool, these results highlight the need for a scalable, context-aware RAG design in industrial TCFL settings.

In the next section, we present our approach in detail by describing the \ourtool pipeline and its components for effective and efficient similarity-based retrieval, which is subsequently leveraged to improve FL performance.

\section{Approach}\label{sec:method}
Building on the work of Saboor et al.~\cite{saboor2025black}, which we refer to as \baseline, we propose \ourtool, a retrieval knowledge generation framework for annotating similar faulty patterns in test scripts for test code fault localization. Unlike \baseline, which performs FL solely based on the failing test case and its associated error message, \ourtool leverages debugging knowledge encoded in a curated corpus of fault-labeled test cases to guide and constrain the reasoning process of large language models (LLMs).

In this setting, we operate under two practical assumptions: (1) a fault-labeled corpus of test cases is available, accumulated over the system's lifetime, where each test failure has been inspected and at least one faulty line has been identified; (2) test failures tend to recur in similar forms and patterns over time.

Fault-labeled test cases corpora naturally arise in continuous integration (CI) environments, version control systems, and issue tracking workflows, where failing tests are repeatedly diagnosed and fixed~\cite{Tomassi2019BugSwarm, Song2022RegMiner, Rene2014Defects4J, Widyasari2020BugsInPy}. Rather than reasoning about a failing test in isolation, \ourtool treats this corpus as an externalized form of debugging memory and reuses it to localize faults in newly failing test cases, also referred to as query tests.

Additionally, test failures often recur in similar forms over time, for example, due to repeated misuse of APIs~\cite{Li2021APImisuse}, fragile assertions~\cite{parry2025systemic}, copy-pasted test logic~\cite{Ain2019codeclone}, or shared environmental assumptions~\cite{Vahabzadeh2015empirical}.
By retrieving fault-related information from test cases similar to the failing query test and explicitly exposing this information to the LLM via targeted annotations, \ourtool reduces the effective FL search space (i.e., the set of all test code elements that could contain the fault) and enables more focused reasoning. Importantly, this is achieved without requiring execution traces, program instrumentation, language-specific analysis, or white-box access to the system under test, making the approach applicable in lightweight and black-box testing scenarios.

When the above assumptions do not hold (e.g., in the presence of entirely novel failure modes or in systems with limited fault-labeled data), \ourtool gracefully degrades to \baseline behavior, maintaining performance at least at the baseline level.

An overview of \ourtool is presented in~\autoref{algorithm1}. 
\begin{algorithm}[tbp]\algosize
\caption{\ourtool: Similar Pattern Annotation via Retrieval Knowledge for Test Code Fault Localization}
\label{algorithm1}
\KwIn{$\tau_f$ --  Query test case}
\KwIn{$D$ --  Fault-labeled test case corpus}
\KwIn{$I$ --  Set of instructions for fault localization}
\KwIn{$k$ --  Number of suspicious lines to return}
\KwOut{$L_{\text{susp}}$ --  Suspicious lines of $\tau_f$, ranked in descending order of suspiciousness}

$D' \gets \textsc{filter}(\tau_f, D, p =  p_{\text{all}})$ \Comment{Definition~\autoref{def::filtering::module}}
~\label{line:filtering}

$S \gets \textsc{search}(\tau_f, D', r=1)$ \Comment{Definition~\autoref{def::similar::engine}}
~\label{line:search}

$X \gets \textsc{retrieve}(S)$ \Comment{Definition~\autoref{def::context::ret}}
~\label{line:retrieve}

$ \tau_f'\gets\textsc{annotate}(\tau_f, X, \epsilon=0.05)$ \Comment{Definition~\autoref{def::annotate}
}
\label{line:annotate}

$L_{\text{susp}} \gets \textsc{TCFL}_\text{base}(\tau_f', I, k)$
\Comment{Definition~\autoref{def::tcfl}}
~\label{line:invoke}

\KwRet $L_{\text{susp}}$

\end{algorithm}
Given a faulty query test case \(\tau_f\), a fault-labeled test case corpus \(D\), a set of FL instructions \(I\), and a desired number of suspicious lines \(k\) to return, \ourtool produces a ranked list of suspicious lines (\(L_{\text{susp}}\)) in \(\tau_f\). The approach consists of four main components: a filtering module (\autoref{line:filtering}), a similarity search engine (\autoref{line:search}), a context retrieval module (\autoref{line:retrieve}), and an annotation module (\autoref{line:annotate}). Together, these components define a fault localization search space, highlighting elements with higher fault likelihood and enabling the LLM to focus on a prioritized subset rather than treating all elements equally. In our setting, we consider lines as the elements to be highlighted. However, the approach is generalizable to elements at both finer and coarser granularity levels. Once the search space of highlighted lines is established, \ourtool invokes the LLM using the same configuration as the \baseline, with the only difference being the test code input, where selected lines are highlighted based on the four components (\autoref{line:invoke}).

\subsection{Input Data}\label{sub::input}
The primary input to \ourtool is a faulty query test case $\tau_f$, whose failure may be observed at any point in the CI pipeline, including development-time testing and operational regression testing. The second input is a fault-labeled test case corpus $D$, where each test case is associated with at least one labeled faulty line. The corpus $D$ is collected across the lifetime of a system and is not necessarily temporally ordered with respect to the query test case. As a result, it may contain test cases whose failures occurred either before or after the failure of $\tau_f$. Such temporal heterogeneity is common in CI pipelines, where test failures are diagnosed, fixed, and documented asynchronously, with bugs triaged based on their severity, priority, and scope~\cite{Hu2014bugtriage}. Temporal availability is explicitly controlled by \ourtool's filtering module, described in~\autoref{sub::filtering}, to support different data availability scenarios and to study their impact on FL effectiveness.

As the CI pipeline evolves, the corpus of fault-labeled test cases naturally grows, allowing \ourtool to continuously accumulate and leverage debugging knowledge.
In our experimental evaluation, we remove duplicated test cases from the corpus using domain-specific knowledge. This preprocessing step ensures that the measured effectiveness of \ourtool reflects its ability to generalize across distinct failure instances, rather than being driven by trivial repetition of identical test cases.

The remaining inputs to \ourtool comprise a set of natural-language FL instructions $I$ and a parameter $k$ that denotes the exact number of suspicious lines to be returned. Rather than being enforced programmatically, this parameter is explicitly encoded in the instruction set $I$, which directs the LLM to produce exactly $k$ candidate faulty lines.

\subsection{Filtering Module}
\label{sub::filtering}

As discussed earlier, the fault-labeled test case corpus \(D\) may grow continuously as part of long-lived CI pipelines. In large-scale industrial settings, such corpora can reach substantial size. For example, one corpus maintained by our industrial partner contains on the order of tens of thousands of test cases prior to curation.\footnote{Due to confidentiality constraints, only a small curated subset of this corpus is available for our experiments.} While large corpora are valuable for capturing accumulated debugging knowledge, considering all fault-labeled test cases during retrieval is impractical and may introduce noise, leading to unnecessary computational overhead and reduced relevance. 

The primary goal of \ourtool's filtering module is therefore to narrow the corpus so that the filtered corpus can serve as an input to the similarity search engine described in~\autoref{sub::sim::engine}, which identifies test cases that are most similar to a given query test case. We refer to the filtered corpus as the \emph{knowledge base}.

Beyond scalability, the filtering module is also designed to address the variation in data availability that naturally arises in CI pipelines. As discussed earlier in this section, the fault-labeled test case corpus may be incomplete or not yet fully populated when a failure occurs. For example, some failures may not yet have been labeled, while others may become available only after the query test case has been diagnosed. Depending on organizational policies, deployment constraints, or evaluation objectives, practitioners may choose to include or exclude such test cases in retrieval and analysis. To reflect these realistic conditions, \ourtool defines multiple filtering policies that simulate different data availability scenarios. Formally, the filtering module is defined as follows.

\begin{definition}[\emph{\textbf{Filtering Module}}]\label{def::filtering::module}
Given a query test case $\tau_f$ and a corpus of fault-labeled test cases \(D\), the filtering module \textsc{filter} selects a subset of test cases from \(D\) based on a filtering policy $p$ as follows:

\[
\textsc{filter}({\tau_f}, D, p) = D', \quad D' \subseteq D 
\]

where $D'$ constitutes the knowledge base used as an input for subsequent similarity-based retrieval, and the policy \(p\) determines which test cases are retained in $D'$. 
\end{definition}

In this paper, we investigate four filtering policies to evaluate the effectiveness of \ourtool across varying forms of fault-labeled data availability. 

These policies are applied to the entire corpus \(D\), generating the following subsets of test cases:

\begin{enumerate}
    \item $p_{\text{all}}$: This policy applies no filtering, generating a knowledge base that contains all test cases in the corpus $D$.
    \item $p_{\text{all-preceding}}$: This policy restricts the corpus to test cases whose failure times occur before that of the query test case $\tau_f$, i.e., before the timestamp of its error message, simulating settings where future-labeled failures are unavailable or intentionally excluded.
    \item $p_{\text{closest-by-time}}$: This policy narrows the corpus to a fixed-size subset of fault-labeled test cases whose failure times are closest to that of \(\tau_f\), based on the intuition that failures occurring within a temporal window are more likely to share similar root causes~\cite{parry2025systemic}.
    \item $p_{\text{closest-time-preceding}}$: This policy combines temporal precedence and temporal locality, selecting a fixed-size subset from $p_{\text{all-preceding}}$, consisting only of failures that occurred before $\tau_f$ while remaining temporally close to it.
    \end{enumerate}

The policies $p_{\text{all-preceding}}$ and $p_{\text{closest-time-preceding}}$ model scenarios in which failures are diagnosed and resolved in the order of their occurrence. In contrast, $p_{\text{all}}$ and $p_{\text{closest-by-time}}$ correspond to systems that employ triaged bug fixing, in which resolution is prioritized by bug severity rather than occurrence time.

Some policies may not be applicable in practice, as some may produce empty or very small subsets of the corpus. For instance, $p_{\text{all-preceding}}$ may be empty or limited in size for newly developed systems. For this reason, the filtering module provides multiple policy options, allowing practitioners to adapt retrieval behavior to data availability, system maturity, and evaluation objectives.

Unless stated otherwise, \autoref{line:filtering} of~\autoref{algorithm1} applies the filtering module with the default value $p_{\text{all}}$.

The set of filtering policies is extensible and can be adapted based on data availability or domain-specific considerations, such as incorporating repair time alongside failure time. Evaluating \ourtool under different filtering policies allows us to systematically study the impact of temporal locality and knowledge base size on the quality of similarity retrieval. We further discuss these policies and their implications in~\autoref{subsub::rq4}.

\subsection{Similarity Search Engine}\label{sub::sim::engine}

This module computes similarity between the query test case \(\tau_f\) and other fault-labeled test cases in the knowledge base \(D'\). Its goal is to identify test cases that exhibit globally similar behavioral and failure characteristics, rather than to localize faults in isolation for each individual query test case. We characterize similarity from two complementary perspectives: dynamic runtime behavior and static code structure. These perspectives are captured using embedding representations of runtime error messages and test code lines, respectively.

Given a knowledge base $D'$, \ourtool retrieves the top-$r$ test cases that are most similar to the query test case $\tau_f$ as follows.

\begin{definition}[\emph{\textbf{Similarity Search Engine}}]\label{def::similar::engine}
Given a knowledge base \(D'\), the similarity search engine \textsc{search} constructs a subset of size \(r\) from \(D'\) consisting of the fault-labeled test cases most similar to the query test case \(\tau_f\). Similarity is computed using a function
\(sim\) over vector embeddings of the concatenation of error messages and test code, as follows:

\[
\begin{aligned}
\textsc{search}(\tau_f, D', r) 
&= \\
\arg\max\nolimits_{tc \in D'}^{(r)} 
   \text{sim}\big(\text{encode}(m' \,\Vert\, L'), 
&\;\text{encode}(m \,\Vert\, L)\big)
\end{aligned}
\]

where \(L'\) and \(L\) denote the entire set of lines respectively in the fault-labeled test case \(tc \in D'\) and the query test case \(\tau_f\), and \(m'\) and \(m\) are their corresponding error messages. The operator \(\Vert\) denotes concatenation, and the function \(encode\) maps the concatenated string to a $d$-dimensional vector embedding. The operator \(\arg\max^{(r)}\) returns the set of the \(r\) test cases in $D'$ with the highest similarity scores with respect to \(\tau_f\).

\end{definition}

In~\autoref{algorithm1}, \autoref{line:search} invokes the similarity search engine to construct \(S\). By default, we use $r=1$ to retrieve the single most similar fault-labeled test case to \(\tau_f\), though the framework naturally generalizes to larger values of $r$. In general, the similarity function \texttt{sim} can be instantiated using any suitable similarity measure. In our implementation, we employ cosine similarity as the similarity metric over vector embeddings and select the most similar test cases via a $k$-nearest neighbors (KNN) procedure.

\subsection{Context Retrieval Module}
Once the set of similar test cases $S$ is constructed, a context retriever module extracts fault-relevant information from the retrieved test cases. Rather than incorporating entire test cases, this module focuses on the faulty lines identified during prior debugging. This design choice ensures that only highly relevant fault-related information is propagated forward, avoiding unnecessary distraction of the LLM with unrelated test code. 

We define the relevant contextual information as the contents of the faulty lines responsible for test failures, as formalized below.

\begin{definition}[\emph{\textbf{Context Retrieval Module}}]\label{def::context::ret}
Given the set \(S\) containing the \(r\) faulty test cases most similar to the query test case \(\tau_f\), the context retrieval module \textsc{retrieve} returns contextual information from the test cases in \(S\) as follows:

\[
\textsc{retrieve}(S) = \bigcup_{i=1}^{r} F_i
\]

where \(F_i\) denotes the set of faulty lines' content in the \(i\)-th test case in \(S\).
\end{definition}

In~\autoref{algorithm1}, \autoref{line:retrieve} computes the set of contents of the faulty lines in the most similar test case.

\subsection{Annotation Module}\label{sub::annotate}
The annotation module uses fault-relevant information retrieved by the context retrieval module to selectively highlight suspicious lines in the query test case \(\tau_f\). The goal of this module is not to directly localize faults, but to guide the LLM's attention toward a small subset of lines that are textually similar to previously observed faulty patterns, i.e., fault locations identified by the context retrieval module. By doing so, \ourtool guides the LLM's reasoning toward the smaller subset while preserving its ability to perform a broader analysis of the entire test case when necessary.

Let \(X\) denote the set of faulty line contents retrieved from the most similar test cases. For each line \(l_i\) in the query test case \(\tau_f\), the annotation module computes a line-level distance score based on its textual distance to the faulty lines in \(X\). Specifically, we define the distance  score of a line \(l_i\) as the minimum of the normalized Levenshtein distances to the faulty lines in \(X\):

\[
w(l_i) = \min_{f \in X} \text{dist}(l_i, f),\; 1\leq i \leq n
\]

where \(\text{dist}(\cdot,\cdot)\) denotes the normalized Levenshtein distance~\cite{levenstein1966} and \(n\) is the total number of lines in \(\tau_f\). This formulation captures the intuition that a line in the query test case is suspicious if it is textually similar to any previously observed faulty line, even if the two lines are not identical.

Using this distance score, the annotation module selectively augments lines in \(\tau_f\) with an explicit annotation message indicating a high likelihood of faultiness. This process is formalized as follows.

\begin{definition}[\emph{\textbf{Annotation Module}}]\label{def::annotate}
Given a query test case \(\tau_f\), a set of retrieved faulty line contents \(X\), and a distance threshold \(\epsilon\), the annotation module selectively augments the lines of \(\tau_f\) as follows:

\[
\textsc{annotate}(\tau_f, X, \epsilon)
=
\{\textsc{tag}(l_i, X, \epsilon) \mid l_i \in L\},
\]

where

\[
\textsc{tag}(l_i, X, \epsilon) =
\begin{cases}
l_i \oplus msg, & \text{if } w(l_i) \leq \epsilon,\\[2mm]
l_i, & \text{otherwise.}
\end{cases}
\]

Here, \(L = \{l_1, l_2, \dots, l_n\}\) denotes the set of lines in \(\tau_f\), \(\oplus\) denotes concatenation, and \(msg\) is an annotation message indicating a high likelihood of faultiness.
\end{definition}

In this paper, we define the annotation message as  
\texttt{"\# !!! high likelihood of being faulty !!!"}. 
The annotation is formatted as a comment-style marker, structurally distinct from both executable code and natural-language instructions. Prior work on prompt-based learning and prompt engineering has shown that the structure and formatting of prompts can influence how LLMs interpret and follow instructions, motivating the use of explicit and clearly delineated input representations~\cite{liu2023pretrain, DBLP:journals/corr/abs-2302-11382}. Inspired by this observation, \ourtool introduces a lightweight inline annotation mechanism to make fault-related cues explicit within the code context.

Specifically, the annotation design follows three principles. First, the comment-style formatting aligns with common coding conventions, allowing the model to interpret the annotation as a human-provided signal rather than executable logic. This design choice is motivated by prior work showing that code comments are a meaningful source of semantic guidance for LLMs in code generation and understanding tasks~\cite{song2024code, imani2025inside, chen2024comments}. Second, the use of visually salient symbols makes annotated lines stand out from regular code and comments, increasing their structural prominence within the prompt. Third, the annotation text explicitly quantifies the likelihood of the suspected fault, providing a clear, unambiguous signal without constraining the model's reasoning.

Together, these design choices make the annotation easy for the LLM to interpret, remain agnostic to the underlying model architecture, and avoid assumptions about internal attention mechanisms. By adopting a comment-style annotation commonly used to convey warnings or developer intent in code, \ourtool provides the LLM with an explicit, human-interpretable indication of increased fault likelihood that complements, rather than replaces, its own reasoning capabilities.

By definition, a line in \(\tau_f\) is annotated whenever it exhibits sufficient textual similarity (i.e., a normalized Levenshtein distance score below the threshold \(\epsilon\)) to at least one previously observed faulty line retrieved from the set of similar test cases. In~\autoref{algorithm1}, \autoref{line:annotate} applies the annotation module using \(\epsilon = 0.05\). We investigate the sensitivity of \ourtool to different threshold values in ~\autoref{rq3}.

Although the annotation mechanism is binary in the current implementation (i.e., a line is either annotated or not), the distance score \(w(l_i)\) naturally enables graded confidence signals, for example, by distinguishing strong from weak matches based on distance values. Such graded annotations could allow the language model to reason over varying degrees of similarity, but we leave this extension to future work. 

Once the annotation process is complete, \ourtool invokes the LLM with the annotated query test case, the FL instructions inherited from \baseline (\autoref{fig:prmpt_template}), where \emph{\{element\}} is set to line, and the desired number of suspicious lines to return. The annotated query test case \(\tau_f'\) consists of the test code lines---some of which are annotated---and the error message. The LLM then produces a ranked list of \(k\) lines that are most likely to be faulty (\autoref{line:invoke}). The ranked list of suspicious lines produced by the LLM can optionally be mapped to coarser granularity levels, such as CFG blocks or functions. This enables FL at multiple abstraction levels, supporting diverse debugging workflows and analysis needs. We discuss this mapping and its impact on evaluation in~\autoref{sec:evaluation}.

\begin{figure*}[!t]
    \centering
    \includegraphics[
        width=\linewidth,
        trim=20 180 330 15,
        clip
    ]{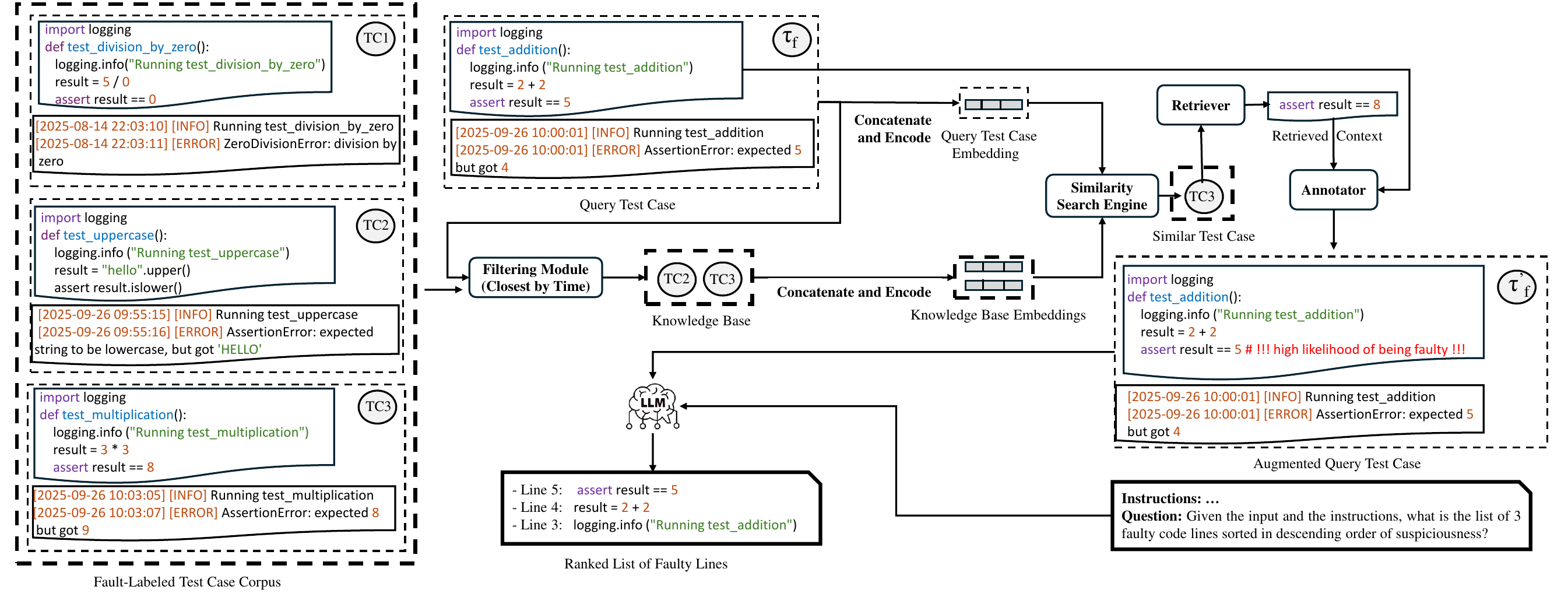}
    \caption{Example workflow of \ourtool.}
    \label{fig:example}
\end{figure*}

\subsection{An Illustrative Example}

To better illustrate our approach,~\autoref{fig:example} presents a demonstrative example consisting of three test cases, denoted as TC1, TC2, and TC3, which together form the fault-labeled test case corpus. Each test case exhibits a distinct failure type: TC1 raises a division-by-zero error, TC2 fails due to an incorrect string case check in an assertion, and TC3 fails because of an incorrect expected result in a multiplication assertion. In the query test case \(\tau_f\), the failure is caused by an incorrect assertion for the addition operation. 

\ourtool begins by filtering the corpus according to a predefined policy to create the knowledge base. In this example, we assume that the filtering policy is $p_{\text{closest-by-time}}$, which retains only those test cases whose failure timestamps are closest to that of the query test case. Based on this policy, TC2 and TC3 are selected because their failures occurred at times similar to the query failure on 2025-09-26.

Next, \ourtool generates vector embeddings for the concatenation of the error message and test code lines of the query test case, as well as for TC2 and TC3 in the knowledge base. Using these embeddings, the similarity search engine identifies TC3 as the most similar test case to the query. Although TC3 involves a multiplication operation, its failure stems from an incorrect numerical assertion that is more semantically similar to the query test case's incorrect addition assertion than to TC2's string-related failure.

The retriever module then extracts the faulty line from TC3 (i.e., \texttt{assert result == 8}) and forwards it to the annotation module together with the lines of the query test case. The annotator module identifies and annotates those lines in the query test case whose content exhibits a small Levenshtein distance to the retrieved faulty line. In this example, \texttt{assert result == 5} in the query test case is textually similar to \texttt{assert result == 8} and is therefore annotated with the message \texttt{"\# !!! high likelihood of being faulty !!!"}.

Finally, the annotated lines, together with the error message and FL instructions, are provided to the LLM, which produces a ranked list of suspicious lines as the final output.

In the following section, we evaluate \ourtool against the \baseline and perform further investigations.

\section{Evaluation}\label{sec:evaluation}

In this section, we evaluate the overall performance of \ourtool. We first introduce the benchmark datasets and describe the experimental design. We then investigate the following research questions to assess the effectiveness, scalability, and robustness of \ourtool across different datasets.

\textbf{RQ1.} To what extent does \ourtool improve fault localization effectiveness compared to \baseline?

\textbf{RQ2.} How does \ourtool compare to the \baseline in terms of scalability and computational efficiency?

\textbf{RQ3.} How sensitive is \ourtool to variations in the annotation threshold?

\textbf{RQ4.} How do different filtering policies affect the performance of \ourtool?

\subsection{Benchmarks}\label{sub::bench}

We conduct our experiments on three datasets, denoted as DS1, DS2, and DS3, which contain 657, 94, and 28 real-world faulty Python test cases, respectively. These datasets were provided by our industrial partner and originate from production testing workflows.

Following prior work~\cite{saboor2025black}, we preprocess each test case by removing blank lines, as they do not contribute to the logic, structure, or behavior of the code and can be safely eliminated to reduce unnecessary complexity~\cite{zhang2024smell,galiullin2024code}. We retain comments, since they have been shown to aid in understanding program intent and code logic~\cite{Steidl2013commnets}.

~\autoref{table::benchmark} summarizes the characteristics of the preprocessed datasets at different levels of code granularity, including the number of files, functions (including constructors and nested functions), control-flow graph (CFG) blocks, abstract syntax tree (AST) statement nodes, and lines of code.

As shown in~\autoref{table::benchmark}, test cases in our datasets range from 56 to 1,429 lines of code. On average, DS2 contains larger test cases than DS1 and DS3, with 395.4 lines per test case. In terms of file structure, test cases in DS1 typically span one or two Python test scripts, whereas those in DS2 and DS3 are each contained within a single script. Structural Characteristics also differ substantially across datasets. DS2 and DS3 contain fewer CFG blocks per test case, averaging 11.3 and 6.0 blocks, respectively, compared to DS1, which averages over 42 CFG blocks per test case. As a result, the ratio of faulty CFG blocks in DS1 is relatively low, averaging 7.8\%, whereas DS2 and DS3 exhibit a higher faulty block ratio of 15.6\% and 21.4\%, respectively. The fewer CFG blocks in DS2 and DS3 suggest that the test cases in these datasets exhibit more sequential control flow, in contrast to the more branching control flow commonly observed in DS1. In addition, DS2 and DS3 contain a substantially larger number of AST statement nodes than CFG blocks (e.g., 202.4 statements on average in DS2, compared to 11.3 CFG blocks), whereas DS1 exhibits a nearly one-to-one correspondence between statements and lines of code (234.8 statements versus 257.2 lines on average). Each test case is accompanied by a log file that records the components invoked by the test codebase, along with test steps that reflect interactions with the SUT's high-level API layer, providing insight into its behavioral complexity through execution paths and cascading interactions. For DS1, the average log file size per test case is 545 KB, ranging from 10 KB to 43,282 KB. The dataset contains an average of 548 test steps per run, with a minimum of 6 and a maximum of 13,935. In DS2, the average log file size is 77 KB, with values ranging from 2 KB to 451 KB, while in DS3 the average is 37 KB, ranging from 12 KB to 66 KB. The average number of test steps is 426 in DS2 (ranging from 2 to 1,823) and 2 in DS3 (ranging from 1 to 11). The test codebase---including supporting modules and functions used by the test scripts---contains 5,159,790 lines for DS1 test cases, and 13,416,809 lines for the test cases in DS2 and DS3.

To identify faulty elements in each test case and establish ground truth for FL, we compute the code \emph{diff} between the preprocessed faulty version of each test case and its corresponding preprocessed repaired version, defined as the first subsequent commit in the repository history in which the failure no longer occurs. New lines introduced only in the repaired version and lacking counterparts in the faulty version are disregarded. Therefore, faulty lines are identified as those removed or modified in the repaired version. Based on the identified faulty lines, we determine the enclosing code statements, blocks, functions, and files, and label them as faulty as well. Across our datasets, most test cases contain multiple faulty locations, which are not necessarily contiguous. On average, faults span 5.97 lines in DS1, 3.56 lines in DS2, and 2.11 lines in DS3. We treat these fault locations as the ground truth, noting that they differ from error locations. Our evaluation supports both single-fault and multi-fault scenarios by treating all modified or removed lines as faulty.

Distinguishing refactoring changes from actual bug fixes in code diffs is non-trivial. To mitigate this issue, we adopt the semi-automated refinement process proposed by Saboor et al.~\cite{saboor2025black}. Specifically, we first identify outlier test cases exhibiting an unusually large number of removed or modified lines of code using the \emph{3-\(\sigma\) rule}~\cite{pukelsheim1994}. Then, this small set of outliers is manually inspected by experts to identify true fault locations and to exclude test cases that involve only refactoring changes. Using this procedure, we excluded 21 test cases across the three datasets, resulting in the final dataset sizes reported in the second column of~\autoref{table::benchmark}.

\begin{table*}[!t]
    \centering
    \scriptsize
    \resizebox{\textwidth}{!}{
    \begin{tabular}{c c l cccc cccc}
        \toprule
        \multirow{2.5}{*}{Dataset} & \multirow{2.5}{*}{\# Test Cases} & \multirow{2.5}{*}{\shortstack{Code \\ Granularity}} & \multicolumn{4}{c}{\# Total} & \multicolumn{4}{c}{Faulty Ratio (\%)} \\ 
        \cmidrule(lr){4-7} \cmidrule(lr){8-11} 
       & & & Min & Median & Mean & Max & Min & Median & Mean & Max \\  
        \midrule
       \multirow{4}{*}{DS1} & \multirow{4}{*}{657} & File  & 1.0  & 2.0 & 1.9 & 2.0 & 50.0 & 50.0 & 54.6 & 100.0  \\  
       & & Function  & 2.0 & 5.0 & 6.0 & 26.0 & 0.0 & 20.0 & 22.6 &  100.0 \\   
       & & Block  & 3.0 & 34.0 & 42.1 & 280.0 & 0.4 & 5.0 & 7.8 & 55.2  \\ 
       & & Statement & 56.0 & 199.0 & 234.8 & 949.0 & 0.1 & 1.0 & 2.7 & 35.7  \\ 
       & & Line  & 56.0 & 211.0 & 257.2 & 1,070.0 & 0.1 & 1.0 & 2.7 & 35.4  \\  
        \midrule
       \multirow{4}{*}{DS2} & \multirow{4}{*}{94} & File  & 1.0 & 1.0 & 1.0 & 1.0 & 100.0 & 100.0 & 100.0 & 100.0  \\  
       & & Function  & 4.0 & 5.0 & 6.6 & 21.0 & 4.8 & 20.0 & 19.3 & 50.0  \\   
       & & Block  & 5.0 & 6.0 & 11.3 & 53.0 & 1.9 & 16.7 & 15.6 & 50.0   \\  
       & & Statement & 54.0 & 178.5 & 202.4 & 479.0 & 0.2 & 0.7 & 1.0 & 9.8 \\
       & & Line  & 64.0 & 279.0 & 395.4 & 1,429.0 & 0.1 & 0.6 & 1.0 & 11.6  \\  
        \midrule
      \multirow{4}{*}{DS3}  & \multirow{4}{*}{28} & File  & 1.0 & 1.0 & 1.0 & 1.0 & 100.0 & 100.0 & 100.0 & 100.0  \\  
       & & Function  & 5.0 & 5.0 & 5.1 & 8.0 & 12.5 & 20.0 & 25.3 & 40.0  \\   
       & & Block  & 6.0 & 6.0 & 6.0 & 6.0 & 16.7 & 16.7 & 21.4 & 33.3  \\   
     & & Statement & 44.0 & 87.0 & 99.3 & 198.0 & 0.5 & 1.6 & 1.9 & 4.8 \\ 
       & & Line  & 83.0 & 167.5 & 195.3 & 501 & 0.2 & 0.9 & 1.2 & 3.6  \\  
        \bottomrule
    \end{tabular}
    }
    \caption{Distribution of total and faulty levels of code granularity in our benchmarks.}
    \label{table::benchmark}
\end{table*}

\subsection{Experimental Design}\label{sub::design}

In this section, we describe the experimental design used to evaluate the effectiveness, efficiency, and robustness of \ourtool in addressing our research questions. To address RQ1 and RQ2, we compare \ourtool against the \baseline~\cite{saboor2025black}, which, to the best of our knowledge, is the only existing LLM-based approach specifically designed for TCFL. \baseline serves as a strong comparison point, as it targets the same problem setting and operates under the same black-box assumption, i.e., it does not require access to the SUT's source code. The two approaches differ fundamentally, however, in how contextual knowledge is constructed and provided to the LLM. We evaluate \ourtool against \baseline by comparing their ability to identify suspicious code elements across multiple datasets and FL granularity levels, thereby enabling a systematic and fair assessment of their relative strengths and limitations.

Since finer-grained FL increases the number of elements that the LLM must reason about, we begin our evaluation at the line level, which represents the finest granularity and is hence the most challenging setting. Following prior work~\cite{saboor2025black}, we configure the number of returned lines ($k$) to 1 (the most suspicious line), 3, 5, and 10. To enable fair and consistent evaluation at coarser granularities, we aggregate line-level FL results into higher-level code structures, such as CFG blocks or AST statements.

As discussed in~\autoref{sub::bench}, DS2 and DS3 contain a substantially larger number of AST statement nodes than CFG blocks per test case, making statement-level FL more challenging than block-level FL for these datasets. In contrast, DS1 exhibits a near one-to-one correspondence between statements and lines of code; therefore, statement-level FL in DS1 provides no substantial additional insight beyond line-level analysis. Based on these observations, to assess FL's performance at a coarser granularity level, we adopt the following strategy. For DS1, we map each identified faulty line to its corresponding CFG block and mark the entire block as faulty. For DS2 and DS3, we map each identified faulty line to its enclosing AST statement node. In the remainder of this paper, we refer to CFG blocks in DS1 and AST statement nodes in DS2 and DS3 as \emph{logical units} or simply \emph{units}, and denote the process of mapping faulty lines to their corresponding units as unit-level FL.

We use \texttt{Qwen2.5-72B}\footnote{https://huggingface.co/Qwen/Qwen2.5-72B} as the underlying LLM for both \ourtool and \baseline across all experimental settings. We choose this model since it is the primary LLM used in the \baseline study, ensuring a fair and controlled comparison. In \ourtool, we use OpenSearch\footnote{https://opensearch.org/} to index the vector representation of fault-labeled test cases. For each test case, we concatenate the error message with the test code lines and generate embeddings using \texttt{bge-m3}\footnote{https://huggingface.co/BAAI/bge-m3}, a general-purpose embedding model with a context window of 8,192 tokens. The resulting embeddings are stored in OpenSearch and queried to support similarity-based retrieval. We configure OpenSearch with an HNSW-based FAISS~\cite{10.1109/TPAMI.2018.2889473,douze2024faiss} backend to enable scalable approximate nearest-neighbor retrieval. To handle long inputs, we adopt a chunk-based embedding strategy implemented via LangChain's recursive character text splitter.\footnote{https://js.langchain.com/docs/concepts/text\_splitters/} We apply a sliding-window overlap of 10\% of \texttt{bge-m3}'s context window (approximately 800 tokens) to preserve contextual continuity across adjacent chunks. All embeddings have dimension 1,024.

To emulate previously unseen query test cases, we follow a leave-one-out evaluation protocol in which each test case is treated in turn as the query. For each query, we remove its ground-truth fault labels and retain fault information for all remaining test cases, thereby forming a corpus of test cases with retained fault information (labels). Filtering the corpus to build the knowledge base, indexing all test cases, generating embeddings, and retrieving similar test cases and their faulty lines incur minimal computational overhead, averaging under two seconds per query test case across all datasets. 

For the experiments addressing RQ1 and RQ2, we adopt the default configuration of \ourtool, as shown in~\autoref{algorithm1}: the filtering module uses the $p_{\text{all}}$ policy (i.e., no filtering), the annotator module employs a distance threshold of 0.05 (\(\epsilon\) in Definition~\autoref{def::annotate}), and the retriever module retrieves one similar test case for each query test case. Later, we evaluate the effect of using annotation thresholds other than 0.05 when addressing RQ3, and we analyze the impact of alternative filtering policies in RQ4.

To address RQ3, we evaluate \ourtool using distance thresholds of 0.15, 0.10, and 0.00 and compare the results with those obtained using the default threshold of 0.05. Larger threshold values allow a greater number of query test lines to be annotated, whereas a threshold of 0.00 annotates a query test case line if it exactly matches one of the retrieved faulty lines. In addition to the FL effectiveness metrics, we report the distribution of the number of annotated lines produced under each threshold setting to better understand their practical impact.

To address RQ4, we evaluate the performance of \ourtool using the four filtering policies defined in~\autoref{sub::filtering}: $p_{\text{all}}$, $p_{\text{all-preceding}}$, $p_{\text{closest-by-time}}$, and $p_{\text{closest-time-preceding}}$. To this end, we run the filtering module once per policy to generate a filtered policy-specific corpus of fault-labeled test cases (i.e., a policy-specific knowledge base). From each knowledge base, the remaining steps follow~\autoref{algorithm1}. Specifically, candidate test cases are compared with the query test case to identify the most similar example, which is then annotated using a distance threshold of 0.05. 

All experiments are conducted on a system equipped with a 72-core CPU, 128 GB of RAM, and four 32GB GPUs. 

\subsection{Evaluation Metrics}\label{sub::metrics}

Similar to prior FL studies~\cite{saboor2025black,DBLP:journals/tse/WenCTWHHC21,DBLP:conf/compsac/LiBWL20}, in addition to Precision@$k$ and Recall@$k$, we evaluate performance using Hit@$k$, MAP@$k$, and MRR@$k$~\cite{manning2008introduction}.

Hit@$k$ measures the proportion of test cases in which at least one faulty element is ranked among the top-$k$ predictions.

MAP@$k$ (Mean Average Precision at $k$) evaluates both the correctness and ranking quality of faulty elements among the top-$k$ results across all test cases. For each test case, the average precision is defined as:

\begin{equation}
\text{AP@}k_i = \frac{1}{m_i} \sum_{j=1}^{k} P_i(j) \cdot \text{rel}_i(j)
\end{equation}

where \(m_i\) is the number of correctly identified faulty elements among the top-\(k\) predictions in the \(i\)th test case. \(P_i(j)\) represents the precision at position \(j\), and \(\text{rel}_i(j)\) is 1 if the element at position \(j\) is faulty, and 0 otherwise. 

MAP@$k$ is then computed as:

\begin{equation}
\text{MAP@}k = \frac{1}{N} \sum_{i=1}^{N} \text{AP@}k_i
\end{equation}

where \(N\) denotes the total number of test cases in the dataset.

MRR@$k$ (Mean Reciprocal Rank at $k$) captures how early the first faulty element appears in the ranked list. For each test case, the reciprocal rank is defined as:

\begin{equation}
RR_i =
\begin{cases}
\frac{1}{\text{rank}_i}, & \text{if a faulty element appears in the top-}k, \\
0, & \text{otherwise}.
\end{cases}
\end{equation}

where \(\text{rank}_i\) denotes the rank of the first correctly identified faulty element in the top-\(k\) results of the \(i\)th test case.

For a dataset of size \(N\), MRR@$k$ is then computed as:

\begin{equation}
\text{MRR@}k = \frac{1}{N} \sum_{i=1}^{N} RR_i.
\end{equation}

Higher values of Hit@$k$, MAP@$k$, and MRR@$k$ indicate better FL performance.

In addition to FL effectiveness, we compare the scalability and efficiency of \ourtool and the \baseline. 
Scalability is measured as the average number of input and output tokens consumed per test case, while efficiency is evaluated by LLM inference time, reported as both the average time per test case and the total time across all test cases for each experiment. 
These metrics capture the computational cost of each approach and are particularly informative for large test suites. Input token counts are computed over the entire LLM prompt, which includes the test code, error messages, annotations, and task instructions, while output token counts reflect the full generated response from the LLM.

\begin{table*}[!t]
    \centering
    \resizebox{\textwidth}{!}{
    \begin{tabular}{c c c cc cc cc cc cc}
        \toprule
        \multirow{2.5}{*}{Dataset} & \multirow{2.5}{*}{Granularity} & \multirow{2.5}{*}{$k$} & \multicolumn{2}{c}{Precision@$k$ (\%)} & \multicolumn{2}{c}{Recall@$k$ (\%)} & \multicolumn{2}{c}{Hit@$k$ (\%)} & \multicolumn{2}{c}{MAP@$k$ (\%)} & \multicolumn{2}{c}{MRR@$k$ (\%)} \\
        \cmidrule(lr){4-5}\cmidrule(lr){6-7}\cmidrule(lr){8-9}\cmidrule(lr){10-11}\cmidrule(lr){12-13}
        & & & Baseline & \ourtool & Baseline & \ourtool & Baseline & \ourtool & Baseline & \ourtool & Baseline & \ourtool\\
        \midrule

        \multirow{8.5}{*}{DS1} & \multirow{4}{*}{Line} & 1 & 44.1 & 54.2  & 23.9 & 30.7 & 44.1 & 54.2  & 45.8 & 55.9 & 45.8 & 55.9 \\
   & & 3 & 26.7 & 39.4 & 33.6 & 50.0 & 56.6 & 74.3 & 51.2 & 65.0 & 51.3 & 65.2 \\ 
     &  & 5 & 21.6 & 31.3 & 39.5 & 55.3 & 63.3 & 77.8 & 52.3 & 64.2 & 53.1 & 64.6\\
     &  & 10 & 15.8 & 21.6  & 46.5  & 62.1  & 68.0 & 80.7 & 51.6 & 61.8 & 53.5  & 63.0 \\
  \cmidrule(lr){2-13}
     & \multirow{4}{*}{Unit} & 1 &  66.8 & 73.7 &       51.7 & 57.2 & 66.8 & 73.7 &     66.8 & 73.7 & 66.8 & 73.7\\
     &  & 3 &  32.9     & 38.6 & 64.7 & 74.1 &  79.5 & 88.7 &   72.5 & 80.0 &   72.4 & 80.1\\
     &  & 5 &  22.6 & 26.8 &    69.4 & 78.1 & 83.0 & 90.4 &     73.7 & 79.2 & 74.0 & 79.4\\
     & & 10 & 13.3 & 15.5 & 74.3 & 81.8 &       85.5 & 91.6 &   73.6 & 77.6 &   74.1 & 78.2 \\

    \midrule
        \multirow{8.5}{*}{DS2} & \multirow{4}{*}{Line} & 1 & 21.3 & 40.4  & 13.7 & 24.6 & 21.3 & 40.4  & 21.3 & 40.4 & 21.3 & 40.4 \\
   & & 3 & 12.8 & 24.1 & 19.5 & 39.2 & 31.9 & 57.4 & 23.8 & 46.8 & 23.6 & 47.5 \\ 
     &  & 5 & 9.1 & 17.0 & 24.6 & 45.0 & 34.0 & 59.6 & 25.7 & 44.9 & 25.6 & 46.0\\
     &  & 10 & 6.3 & 9.7  & 32.7  & 47.5  & 43.6 & 63.8 & 26.3 & 43.0 & 27.4  & 45.0 \\
  \cmidrule(lr){2-13}
     & \multirow{4}{*}{Unit} & 1 &  44.7 & 46.8 &       33.6 & 35.5 & 44.7 & 46.8 &     44.7 & 46.8 & 44.7 & 46.8\\
     &  & 3 &  18.4     & 26.6 & 39.3 & 54.5 &  53.2 & 67.0 &   47.2 & 56.3 &   47.2 & 56.4\\
     &  & 5 &  12.3 & 17.2 &    43.0 & 57.3 & 55.3 & 67.0 &     49.2 & 54.1 & 49.5 & 55.0\\
     & & 10 & 7.3 & 8.3 & 48.1 & 54.6 & 62.8 & 69.1 &   50.0 & 55.5 &   51.0 & 56.2 \\

    \midrule
        \multirow{8.5}{*}{DS3} & \multirow{4}{*}{Line} & 1 & 42.9 & 57.1  & 34.2 & 46.1 & 42.9 & 57.1  & 42.9 & 57.1 & 42.9 & 57.1 \\
   & & 3 & 21.4 & 34.5 & 52.3 & 66.2 & 60.7 & 85.7 & 51.2 & 74.4 & 51.2 & 74.4 \\ 
     &  & 5 & 13.6 & 22.1 & 50.8 & 66.2 & 60.7 & 85.7 & 49.5 & 69.9 & 49.8 & 69.9\\
     &  & 10 & 6.8 & 12.9  & 52.6  & 74.5  & 60.7 & 89.3 & 48.5 & 68.1 & 50.0  & 69.2 \\
  \cmidrule(lr){2-13}
     & \multirow{4}{*}{Unit} & 1 &  71.4 & 82.1 &       63.7 & 72.0 & 71.4 & 82.1 &     71.4 & 82.1 & 71.4 & 82.1\\
     &  & 3 &  25.0     & 44.0 & 65.5 & 86.6 &  75.0 & 100.0 &  73.2 & 97.6 &   73.2 & 98.2\\
     &  & 5 &  15.0 & 27.9 &    65.5 & 88.1 & 75.0 & 100.0 &    73.2 & 94.9 & 73.2 & 94.6\\
     & & 10 & 7.9 & 15.7 & 65.8 & 95.2 &        78.6 & 100.0 &  72.6 & 93.8 &   72.6 & 94.6 \\

        \bottomrule
    \end{tabular}

    }
    \caption{The fault localization effectiveness of \ourtool compared to \baseline using Qwen2.5:72B, evaluated at line and unit levels of granularity for different numbers of requested faulty lines ($k$ values) across the three datasets.}
    \label{table::accuracy::results}
\end{table*}

\subsection{Results}\label{sub::results}
\subsubsection{RQ1: To what extent does \ourtool improve fault localization effectiveness compared to \baseline? }\label{subsub::rq1}

~\autoref{table::accuracy::results} compares the FL performance of \ourtool and \baseline across three datasets, at both line and unit levels of granularity, and multiple values of $k$.

Across all datasets and evaluation settings, \ourtool consistently outperforms the \baseline in terms of FL effectiveness. In DS1, \ourtool achieves higher scores across all metrics at both line and unit granularities. For instance, at line-level FL, Precision@3 improves from 26.7\% with the \baseline to 39.4\% with \ourtool, corresponding to an improvement of nearly 13 percentage points (pp), or over 47\%. Precision improvements persist across other $k$ values, with gains of 10.1 pp, 9.7 pp, and 5.8 pp for top-1, top-5, and top-10 line-level FL, respectively. Similar trends are observed for Recall@$k$ and Hit@$k$, which reach 81.8\% and 91.6\%, respectively, for top-10 unit-level FL. MAP@$k$ and MRR@$k$ also show consistent improvements, with MAP@$k$ increasing by up to 13.8 pp at top-3 line-level and by at least 4 pp at top-10 unit-level. 

In DS2, which contains larger test cases, \ourtool yields even more pronounced improvements. Specifically, line-level Precision@1 increases by 19.1 pp, while Precision@3 improves by 11.3 pp compared to \baseline. Recall@$k$ and Hit@$k$ also exhibit significant enhancements, with line-level Hit@3 increasing from 31.9\% for the \baseline to 57.4\% for \ourtool. A similar pattern of improvement is observed in DS3, where \ourtool consistently achieves higher effectiveness across all metrics. Notably, unit-level Hit@$k$ reaches 100\% for \(k=\)3, 5, and 10, indicating that \ourtool successfully identifies at least one faulty unit among the top-ranked results for all test cases under these settings.

\Finding{Providing the LLM with annotated patterns of previously observed faulty test code lines, derived from similar retrieved test cases, leads to significant improvements in FL performance. Specifically, at the line level, \ourtool improves Precision and Hit by 10.1--19.1 percentage points (pp) and Recall by 6.8--11.9 pp for top-1 localization. At top-10 line-level FL, Precision increases by 3.4--6.1 pp, Recall by 14.8--21.9 pp, and Hit by 12.7--28.6 pp. These gains are consistently observed across all datasets, evaluation effectiveness metrics, and granularity levels, highlighting the effectiveness of retrieval pattern annotation in guiding LLM-based fault localization.}

\subsubsection{RQ2: How does \ourtool compare to the \baseline in terms of scalability and computational efficiency?}\label{subsub::rq2}

Scalability and computational efficiency results are presented in~\autoref{table::efficiency::results}. As mentioned in~\autoref{sub::design}, FL is performed at the line level and faulty lines are subsequently mapped to their corresponding units in an offline post-processing step. Therefore, the reported metrics in ~\autoref{table::efficiency::results} reflect only the cost of line-level FL.

Generally, for both \ourtool and \baseline, input token usage increases approximately linearly with test case size due to the larger amount of test code included in the prompt, whereas output token usage increases with larger $k$, as the LLM must generate more candidate faulty locations. Compared to the \baseline, \ourtool incurs prompts with a small increase in average input tokens due to the inclusion of annotation messages appended to selected test code lines. This overhead is modest, averaging approximately 50--60 additional input tokens per test case for DS1, 20--30 for DS2, and 10--20 for DS3. A similarly small increase is observed in the average number of output tokens in most cases, since annotation messages are occasionally preserved in the LLM's generated responses.

Regarding inference time, \ourtool performs comparably to the \baseline across all datasets. For DS1 and DS2, the average inference time per test case remains effectively unchanged. This trend also largely holds for DS3, with only a negligible increase of less than 4 seconds per test case at \(k=\)3 and 5. Overall, these results indicate that the additional annotation and retrieval steps introduced by \ourtool do not impose a significant computational burden. From a practical perspective, these overheads remain sufficiently small and comparable to the baseline, whose execution cost is already manageable for CI environments. This suggests that \ourtool can be integrated into CI-based debugging workflows without substantially impacting debugging turnaround time or analysis scalability.

\Finding{\ourtool introduces only a minor computational overhead, requiring approximately 10--60 additional input tokens per test case across all datasets and experimental configurations. This results in only a negligible increase in output token and inference time, adding at most 3.9 seconds per test case compared to the baseline. When considered alongside the substantial effectiveness gains observed in RQ1, these results indicate that \ourtool achieves improved fault localization performance while maintaining scalability and efficiency comparable to \baseline.}

\begin{table*}[!t]
    \centering
    \tiny
    \resizebox{\textwidth}{!}{
    \begin{tabular}{c c cccc cccc}
        \toprule
        \multirow{4.5}{*}{Dataset}  & \multirow{4.5}{*}{$k$} & \multicolumn{4}{c}{Average Token (\#)} & \multicolumn{4}{c}{Inference Time} \\
        \cmidrule(lr){3-6}\cmidrule(lr){7-10}
        & & \multicolumn{2}{c}{Baseline} & \multicolumn{2}{c}{\ourtool} & \multicolumn{2}{c}{Baseline} & \multicolumn{2}{c}{\ourtool}\\
         \cmidrule(lr){3-4}\cmidrule(lr){5-6}
         \cmidrule(lr){7-8}\cmidrule(lr){9-10}
        & & In & Out & In & Out & Avg (sec) & Sum & Avg (sec) & Sum \\
        \midrule

        \multirow{4}{*}{DS1}  & 1 & 6.50k & 24 & 6.56k & 26 & 38.0 & 6h 56m 8s & 38.3 & 6h 59m 16s \\
    & 3 & 6.50k & 69 & 6.56k & 71 & 41.3 & 7h 32m 34s & 41.6 & 7h 35m 59s\\ 
       & 5 & 6.50k & 112 & 6.56k & 116 & 44.3 & 8h 5m 2s & 44.9 & 8h 11m 23s\\
     & 10 & 6.51k & 216 & 6.56k & 224 & 52.2 & 9h 31m 4s & 52.8 & 9h 38m 37s\\
     \midrule

    \multirow{4}{*}{DS2}  & 1 & 9.27k & 30 & 9.30k & 29 & 46.6 & 1h 12m 55s & 46.5 & 1h 12m 54s \\
    & 3 & 9.27k & 86 & 9.30k & 80 & 50.5 & 1h 19m 3s & 50.1 & 1h 18m 32s\\ 
       & 5 & 9.27k & 128 & 9.30k & 130 & 53.6 & 1h 23m 56s & 53.8 & 1h 24m 21s\\
     & 10 & 9.28k & 235 & 9.30k & 244 & 61.4 & 1h 36m 12s & 62.2 & 1h 37m 22s\\
     \midrule

    \multirow{4}{*}{DS3}  & 1 & 5.44k & 53 & 5.45k & 53 & 37.5 & 16m 16s & 37.5 & 16m 15s \\
    & 3 & 5.44k & 138 & 5.45k & 194 & 43.4 & 18m 49s & 47.3 & 20m 29s\\ 
       & 5 & 5.44k & 252 & 5.45k & 284 & 51.7 & 22m 25s & 54.1 & 23m 25s\\
     & 10 & 5.44k & 417 & 5.46k & 409 & 63.2 & 27m 23s & 62.6 & 27m 8s\\

        \bottomrule
    \end{tabular}

    }
    \caption{Average input (In) and output (Out) token counts, along with total and average inference time per test case for \ourtool compared to \baseline using Qwen2.5:72B, evaluated at line level of granularity for different numbers of requested faulty lines ($k$) across the three datasets.}
    \label{table::efficiency::results}
\end{table*}

\begin{table*}[!t]
\centering
\resizebox{\linewidth}{!}{
\begin{tabular}{c 
                cccc 
                cccc  
                cccc 
                cccc}
\toprule
\multirow{2.5}{*}{Dataset} & 
\multicolumn{4}{c}{$\epsilon$ = 0.15} &
\multicolumn{4}{c}{$\epsilon$ = 0.10} &
\multicolumn{4}{c}{$\epsilon$ = 0.05} &
\multicolumn{4}{c}{$\epsilon$ = 0.00} \\

\cmidrule(lr){2-5} 
\cmidrule(lr){6-9}
\cmidrule(lr){10-13} 
\cmidrule(lr){14-17}

& Min & Mean & Median & Max &
Min & Mean & Median & Max &
Min & Mean & Median & Max &
Min & Mean & Median & Max \\

\midrule

DS1 & 0 (0.0\%) & 7.3 (3.1\%) & 2 (1.0\%) & 102 (45.3\%) & 0 (0.0\%) & 6.6 (2.8\%) & 2 (0.9\%) & 102 (44.5\%) & 0 (0.0\%) & 6.1 (2.6\%) & 2 (0.7\%) & 100 (44.5\%) & 0 (0.0\%) & 5.7 (2.4\%) & 1 (0.6\%) & 95 (43.1\%) \\
\midrule 
DS2 & 0 (0.0\%) & 4.1 (1.3\%) & 1 (0.5\%) & 43 (6.4\%) & 0 (0.0\%) & 3.8 (1.2\%) & 1 (0.4\%) & 39 (6.4\%) & 0 (0.0\%) & 3.2 (1.1\%) & 1 (0.3\%) & 36 (5.8\%) & 0 (0.0\%) & 1.6 (0.5\%) & 0 (0.0\%) & 34 (4.7\%)\\
\midrule 
DS3 & 0 (0.0\%) & 1.4 (0.9\%) & 1 (0.3\%) & 4 (2.9\%) & 0 (0.0\%) & 1.2 (0.7\%) & 0 (0.0\%) & 4 (2.9\%) &  0 (0.0\%) & 1.2 (0.7\%) &  0 (0.0\%) & 4 (2.9\%) & 0 (0.0\%)  & 0.5 (0.3\%) & 0 (0.0\%)  & 3 (1.8\%)\\
\bottomrule

\end{tabular}
}
\caption{Distribution of the number of annotated test case lines and their corresponding ratios to the total number of lines for distance threshold (\(\epsilon\)) values of 0.15, 0.10, 0.05, and 0.00 across three datasets.}
\label{table::annotation::count}
\end{table*}

\begin{table*}[!t]
\centering
\resizebox{\textwidth}{!}{
\begin{tabular}{c c c
                 cccc   
                 cccc   
                 cccc   
                 cccc   
                 cccc } 
\toprule
\multirow{2.5}{*}{Dataset} & 
\multirow{2.5}{*}{Granularity} & 
\multirow{2.5}{*}{$k$} &
\multicolumn{4}{c}{Precision@$k$ (\%)} & 
\multicolumn{4}{c}{Recall@$k$ (\%)} & 
\multicolumn{4}{c}{Hit@$k$ (\%)} & 
\multicolumn{4}{c}{MAP@$k$ (\%)} & 
\multicolumn{4}{c}{MRR@$k$ (\%)} \\
\cmidrule(lr){4-7} \cmidrule(lr){8-11} \cmidrule(lr){12-15} \cmidrule(lr){16-19} \cmidrule(lr){20-23}
& & & 0.15 & 0.10 & 0.05 & 0.00 & 0.15 & 0.10 & 0.05 & 0.00 & 0.15 & 0.10 & 0.05 & 0.00 & 0.15 & 0.10 & 0.05 & 0.00 & 0.15 & 0.10 & 0.05 & 0.00 \\
\midrule

\multirow{8}{*}{DS1}
& \multirow{4}{*}{Line}
& 1  & 54.0 & 53.4 & 54.2 & 54.6 
     & 30.5 & 30.2 & 30.7 & 30.6 
     & 54.0 & 53.4 & 54.2 & 54.6 
     & 55.6 & 54.9 & 55.9 & 56.3
     & 55.6 & 54.9 & 55.9 & 56.3 \\
&  & 3  & 38.0 & 39.2 & 39.4 & 39.6 
     & 48.0 & 49.7 & 50.0 & 50.0 
     & 73.4 & 73.5 & 74.3 & 73.7 
     & 63.8 & 64.7 & 65.0 & 64.8
     & 63.9 & 64.9 & 65.2 & 64.6 \\
&  & 5  & 30.6 & 31.3 & 31.3 & 30.5 
     & 54.6 & 56.0 & 55.3 & 54.2 
     & 77.6 & 77.5 & 77.8 & 75.3 
     & 63.5 & 63.0 & 64.2 & 63.2
     & 63.9 & 63.6 & 64.6 & 63.0 \\
&  & 10 & 21.3 & 21.7 & 21.6 & 21.4 
     & 62.5 & 63.6 & 62.1 & 63.4 
     & 81.3 & 81.3 & 80.7 & 81.4 
     & 61.6 & 60.6 & 61.8 & 62.7
     & 62.6 & 62.3 & 63.0 & 63.3 \\
\cmidrule(lr){2-23}

& \multirow{4}{*}{Unit}
& 1  & 72.8 & 72.5 & 73.7 & 73.2 
     & 56.4 & 56.2 & 57.2 & 56.6 
     & 72.8 & 72.5 & 73.7 & 73.2 
     & 72.8 & 72.5 & 73.7 & 73.2
     & 72.8 & 72.5 & 73.7 & 73.2 \\
&  & 3  & 37.6 & 37.8 & 38.6 & 38.1 
     & 71.9 & 72.1 & 74.1 & 73.6 
     & 87.4 & 87.1 & 88.7 & 88.0 
     & 79.0 & 79.3 & 80.0 & 79.6
     & 79.0 & 79.2 & 80.1 & 79.6 \\
&  & 5  & 26.3 & 26.3 & 26.8 & 26.2 
     & 76.8 & 77.4 & 78.1 & 76.7 
     & 89.6 & 90.1 & 90.4 & 89.3 
     & 78.5 & 78.9 & 79.2 & 78.6
     & 78.6 & 79.1 & 79.4 & 78.6 \\
&  & 10 & 15.6 & 15.3 & 15.5 & 15.0 
     & 81.9 & 82.7 & 81.8 & 81.6 
     & 91.8 & 93.0 & 91.6 & 91.9 
     & 78.2 & 77.9 & 77.6 & 78.4
     & 78.4 & 78.5 & 78.2 & 78.8 \\
     \midrule
\multirow{8}{*}{DS2}
& \multirow{4}{*}{Line}
& 1  & 41.5 & 40.4 & 40.4 & 39.4 
     & 24.9 & 24.6 & 24.6 & 22.6 
     & 41.5 & 40.4 & 40.4 & 39.4 
     & 41.5 & 40.4 & 40.4 & 39.4
     & 41.5 & 40.4 & 40.4 & 39.4 \\
&  & 3  & 23.0 & 22.3 & 24.1 & 22.3 
     & 39.0 & 38.1 & 39.2 & 34.2 
     & 57.4 & 57.4 & 57.4 & 51.1 
     & 46.6 & 47.8 & 46.8 & 41.3
     & 47.2 & 47.9 & 47.5 & 41.7 \\
&  & 5  & 17.2 & 17.2 & 17.0 & 16.6 
     & 47.6 & 46.2 & 45.0 & 42.5 
     & 61.7 & 59.6 & 59.6 & 55.3 
     & 46.4 & 45.5 & 44.9 & 42.0
     & 46.8 & 46.7 & 46.0 & 42.2 \\
&  & 10 & 9.7 & 9.7 & 9.7 & 9.3 
     & 49.9 & 48.9 & 47.5 & 45.9 
     & 63.8 & 64.9 & 63.8 & 61.7 
     & 41.8 & 42.6 & 43.0 & 40.6
     & 43.6 & 45.2 & 45.0 & 41.1 \\
\cmidrule(lr){2-23}

& \multirow{4}{*}{Unit}
& 1  & 46.8 & 46.8 & 46.8 & 46.8 
     & 35.5 & 35.5 & 35.5 & 35.5 
     & 46.8 & 46.8 & 46.8 & 46.8 
     & 46.8 & 46.8 & 46.8 & 46.8
     & 46.8 & 46.8 & 46.8 & 46.8 \\
&  & 3  & 26.2 & 25.5 & 26.6 & 25.2 
     & 54.7 & 53.8 & 54.5 & 51.5 
     & 64.9 & 64.9 & 67.0 & 62.8 
     & 55.4 & 55.1 & 56.3 & 53.7
     & 55.5 & 55.0 & 56.4 & 53.9 \\
&  & 5  & 16.8 & 16.4 & 17.2 & 16.8 
     & 57.7 & 55.9 & 57.3 & 56.1 
     & 66.0 & 66.0 & 67.0 & 66.0 
     & 54.5 & 55.4 & 54.1 & 54.8
     & 55.4 & 55.9 & 55.0 & 55.1 \\
&  & 10 & 9.0 & 9.4 & 8.3 & 8.4 
     & 58.2 & 62.2 & 54.6 & 55.8 
     & 68.1 & 72.3 & 69.1 & 67.0 
     & 53.3 & 55.2 & 55.5 & 53.4
     & 54.9 & 56.7 & 56.2 & 54.1 \\
     \midrule
\multirow{8}{*}{DS3}
& \multirow{4}{*}{Line}
& 1  & 60.7 & 50.0 & 57.1 & 57.1 
     & 47.3 & 41.3 & 46.1 & 43.7 
     & 60.7 & 50.0 & 57.1 & 57.1 
     & 60.7 & 50.0 & 57.1 & 57.1
     & 60.7 & 50.0 & 57.1 & 57.1 \\
&  & 3  & 33.3 & 28.6 & 34.5 & 29.8 
     & 64.8 & 57.9 & 66.2 & 62.1 
     & 85.7 & 75.0 & 85.7 & 78.6 
     & 72.0 & 69.0 & 74.4 & 64.6
     & 71.4 & 69.6 & 74.4 & 64.9 \\
&  & 5  & 24.3 & 20.7 & 22.1 & 20.7 
     & 74.0 & 69.9 & 66.2 & 68.3 
     & 89.3 & 89.3 & 85.7 & 89.3 
     & 70.5 & 69.4 & 69.9 & 65.5
     & 73.3 & 69.3 & 69.9 & 65.6 \\
&  & 10 & 10.0 & 14.3 & 12.9 & 11.1 
     & 69.6 & 80.4 & 74.5 & 72.5 
     & 89.3 & 92.9 & 89.3 & 89.3 
     & 67.6 & 73.9 & 68.1 & 65.5
     & 68.9 & 76.2 & 69.2 & 65.6 \\
\cmidrule(lr){2-23}

& \multirow{4}{*}{Unit}
& 1  & 85.7 & 75.0 & 82.1 & 85.7 
     & 73.2 & 67.3 & 72.0 & 73.2 
     & 85.7 & 75.0 & 82.1 & 85.7 
     & 85.7 & 75.0 & 82.1 & 85.7
     & 85.7 & 75.0 & 82.1 & 85.7 \\
&  & 3  & 42.9 & 42.9 & 44.0 & 40.5 
     & 86.3 & 85.7 & 86.6 & 83.6 
     & 100.0 & 96.4 & 100.0 & 100.0 
     & 96.1 & 94.9 & 97.6 & 92.9
     & 96.4 & 94.6 & 98.2 & 92.9 \\
&  & 5  & 30.7 & 28.6 & 27.9 & 25.7 
     & 94.0 & 89.3 & 88.1 & 85.4 
     & 100.0 & 100.0 & 100.0 & 100.0 
     & 91.4 & 93.7 & 94.9 & 92.7
     & 94.6 & 93.8 & 94.6 & 92.9 \\
&  & 10 & 13.9 & 14.6 & 15.7 & 13.6 
     & 89.0 & 91.1 & 95.2 & 87.2 
     & 100.0 & 100.0 & 100.0 & 100.0 
     & 94.1 & 95.1 & 93.8 & 89.8
     & 96.4 & 96.4 & 94.6 & 88.9 \\
     \bottomrule
\end{tabular}
}
\caption{The fault localization effectiveness of \ourtool at line and unit granularity levels across annotation thresholds of 0.15, 0.10, 0.05, and 0.00 on datasets DS1, DS2, and DS3.}
\label{table::annotator_threshold_results}
\end{table*}

\begin{table*}[!t]
\centering
\resizebox{\linewidth}{!}{
\begin{tabular}{cc 
                cc cc cc cc        
                cc cc cc cc }
\toprule

\multirow{4.5}{*}{Dataset} & 
\multirow{4.5}{*}{$k$} &

\multicolumn{8}{c}{Average Token Count} &
\multicolumn{8}{c}{Inference Time} \\

\cmidrule(lr){3-10} 
\cmidrule(lr){11-18}

& &
\multicolumn{2}{c}{0.15} &
\multicolumn{2}{c}{0.10} &
\multicolumn{2}{c}{0.05} &
\multicolumn{2}{c}{0.00} &
\multicolumn{2}{c}{0.15} &
\multicolumn{2}{c}{0.10} &
\multicolumn{2}{c}{0.05} &
\multicolumn{2}{c}{0.00} \\

\cmidrule(lr){3-4} \cmidrule(lr){5-6} \cmidrule(lr){7-8} \cmidrule(lr){9-10}
\cmidrule(lr){11-12} \cmidrule(lr){13-14} \cmidrule(lr){15-16} \cmidrule(lr){17-18}

& &
In & Out &
In & Out &
In & Out &
In & Out &
Avg (sec) & Sum &
Avg (sec) & Sum &
Avg (sec) & Sum &
Avg (sec) & Sum \\

\midrule

\multirow{4}{*}{DS1}
& 1  & 6.57k & 26 & 6.56k & 26 & 6.56k & 26 & 6.56k & 26 
     & 38.4 & 7h 0m 24s & 38.4 & 7h 0m 12s & 38.3 & 6h 59m 16s & 38.2 & 6h 58m 44s \\
& 3  & 6.57k & 72 & 6.56k & 71 & 6.56k & 71 & 6.56k & 72 
     & 41.7 & 7h 36m 53s & 41.7 & 7h 36m 4s & 41.6 & 7h 35m 59s & 41.7 & 7h 36m 8s \\
& 5  & 6.57k & 116 & 6.56k & 116 & 6.56k & 116 & 6.56k & 116 
     & 44.9 & 8h 11m 21s & 44.9 & 8h 12m 6s & 44.9 & 8h 11m 23s & 44.9 & 8h 11m 13s \\
& 10 & 6.57k & 226 & 6.57k & 224 & 6.56k & 224 & 6.56k & 222 
     & 53.1 & 9h 41m 13s & 52.9 & 9h 39m 0s & 52.8 & 9h 38m 37s & 52.8 & 9h 38m 10s \\
\midrule
\multirow{4}{*}{DS2}
& 1  & 9.31k & 30 & 9.31k & 30 & 9.30k & 29 & 9.29k & 29 
     & 46.8 & 1h 13m 21s & 46.5 & 1h 12m 52s & 46.5 & 1h 12m 54s & 46.6 & 1h 12m 55s \\
& 3  & 9.31k & 77 & 9.31k & 78 & 9.30k & 80 & 9.29k & 79 
     & 50.1 & 1h 18m 25s & 50.1 & 1h 18m 32s & 50.1 & 1h 18m 32s & 50.2 & 1h 18m 37s \\
& 5  & 9.31k & 130 & 9.31k & 143 & 9.30k & 130 & 9.29k & 127 
     & 53.9 & 1h 24m 25s & 55.1 & 1h 26m 15s & 53.8 & 1h 24m 21s & 53.6 & 1h 23m 58s \\
& 10 & 9.31k & 246 & 9.31k & 258 & 9.30k & 244 & 9.29k & 258 
     & 62.5 & 1h 37m 57s & 63.4 & 1h 39m 17s & 62.2 & 1h 37m 22s & 63.3 & 1h 39m 7s \\
\midrule
\multirow{4}{*}{DS3}
& 1  & 5.46k & 57 & 5.45k & 54 & 5.45k & 53 & 5.45k & 50 
     & 37.7 & 16m 19s & 37.5 & 16m 14s & 37.5 & 16m 15s & 37.2 & 16m 6s \\
& 3  & 5.46k & 175 & 5.45k & 193 & 5.45k & 194 & 5.45k & 148 
     & 46.3 & 20m 4s & 47.3 & 20m 28s & 47.3 & 20m 29s & 44.2 & 19m 10s \\
& 5  & 5.46k & 244 & 5.45k & 252 & 5.45k & 284 & 5.45k & 242 
     & 51.0 & 22m 4s & 51.7 & 22m 24s & 54.1 & 23m 25s & 51.1 & 22m 7s \\
& 10 & 5.46k & 415 & 5.46k & 427 & 5.46k & 409 & 5.45k & 432 
     & 63.1 & 27m 19s & 63.9 & 27m 41s & 62.6 & 27m 8s & 64.2 & 27m 50s \\
\bottomrule

\end{tabular}
}
\caption{Average input (In) and output (Out) token counts, along with total and average inference time per test case, for line-level FL using \ourtool across datasets with thresholds ranging from 0.00 to 0.15. 
}
\label{table::annotator_threshold_efficiency}
\end{table*}

\subsubsection{RQ3: How sensitive is \ourtool to variations in the
annotation threshold?}\label{rq3}

A key parameter that may influence the behavior of \ourtool is the distance threshold \(\epsilon\) used by the annotator module. In RQ1 and RQ2, we fixed this threshold value to \(\epsilon = 0.05\).  In RQ3, we investigate the sensitivity of \ourtool to variations in this parameter to assess the approach's robustness and dependence on threshold tuning.
Specifically, we investigate how varying degrees of similarity within a meaningful range between the query test case lines and the retrieved faulty lines affect the provided guidance to the LLM for effective FL in our datasets.

As discussed in the experimental design, a larger threshold potentially allows more lines in the query test case to be annotated, whereas a threshold of \(\epsilon = 0.00\) annotates a line from the query test case only when it exactly matches one of the retrieved faulty lines. ~\autoref{table::annotation::count} reports the distribution of the number of annotated lines for threshold values 0.0, 0.05, 0.10, and 0.15 across all datasets, along with the ratios of annotated lines to the total number of test code lines. As expected, decreasing the threshold value from 0.15 to 0.0 reduces the average number of annotated lines (e.g., an average of 7.3 annotated lines per test case in DS1 at \(\epsilon\) = 0.15 compared to 5.7 lines at \(\epsilon\) = 0.0). DS1 consistently exhibits the highest annotation ratios, suggesting greater similarity between the query and retrieved faulty lines, whereas DS3 has the fewest annotated lines, indicating that its test cases are less similar to one another.

In addition to measuring the number of annotated lines across varying thresholds, we assess the impact of threshold variation on \ourtool's performance. Using the same experimental design as RQ1 and RQ2, we vary the annotation threshold \(\epsilon\) in the range 0.00--0.15 and report the results in ~\autoref{table::annotator_threshold_results}. Overall, \ourtool's performance differences across thresholds are modest. Across datasets and evaluation metrics, variations remain largely negligible. This indicates that, to provide impactful contextual guidance to the LLM, strict similarity between the query test and previously observed faulty lines is not required. Consequently, \ourtool is largely insensitive to the precise choice of annotation threshold within a reasonable similarity range that captures meaningful patterns. This demonstrates that valuable contextual guidance to the LLM can still be provided even when retrieved faulty lines are not uniquely similar.   

DS1 exhibits the most consistent results across thresholds at both line and unit granularities, whereas DS2 shows slightly greater variation, and DS3 is the most sensitive. However, the performance trends are not monotonic with respect to the threshold value within the range of 0.00--0.15. For instance, in DS3, the unit-level FL Precision@$k$ decreases from 85.7\% to 82.1\% for \(k=\)1 as the threshold increases from 0.00 to 0.05, but increases from 40.5\% to 44.0\% for \(k=\)3. These observations suggest that neither overly restrictive nor overly permissive thresholds are universally optimal.

To assess the scalability and efficiency implications, we also measured the average number of input/output tokens used and inference times across DS1--DS3 for different threshold values.
~\autoref{table::annotator_threshold_efficiency} summarizes computational efficiency at the line-level. Overall, variation in the annotation threshold has a negligible effect on token usage and inference time. Consistent with RQ2, larger datasets and higher $k$ values increase token counts and inference times, driven by longer prompts and additional model computation. While stricter thresholds may slightly reduce the number of annotations and thus marginally decrease the input token count, the reduction is small (typically 10--20 tokens) and does not meaningfully affect overall scalability. This is expected, since input tokens primarily depend on the query inputs and on the unchanging FL instructions across different $k$ values and datasets. 

Output token counts and inference times exhibit similarly minor fluctuations across thresholds, indicating that threshold settings do not meaningfully affect the length of model-generated outputs or processing time. Importantly, both stricter and more lenient annotation thresholds do not introduce efficiency penalties, suggesting that precise threshold selection can focus solely on effectiveness without compromising computational cost.

Taken together, these results indicate that threshold variation within the range of 0.00--0.15 has minimal impact on FL effectiveness, and practitioners need not tune this hyperparameter aggressively. An extremely strict threshold (\(\epsilon = 0.00\)) risks omitting useful annotations, while highly permissive thresholds (\(\epsilon = 0.15\)) may occasionally introduce less informative annotations. Based on our empirical observations, thresholds in the range of 0.05--0.10 strike a favorable balance between precision and contextual relevance.

\Finding{\ourtool delivers stable, strong FL performance across a wide, reasonable range of annotation thresholds, substantially reducing the need for threshold tuning in practical settings.}

\subsubsection{RQ4: How do different filtering policies affect the performance of  \ourtool?}\label{subsub::rq4}
In this research question, we further examine the performance of \ourtool by evaluating it under the four filtering policies defined in~\autoref{sub::filtering} and by simulating different data availability conditions in CI pipelines. 
Among these policies, $p_{\text{all}}$ applies no filtering, ensuring that all potentially useful examples are available in the knowledge base. However, this may lead to a large corpus, which can increase computational overhead and data noise, thereby reducing retrieval efficiency and effectiveness. The other policies apply filtering based on the failure timestamps extracted from test execution logs. Leveraging the failure time not only reduces the corpus size but may also improve retrieval reliability by focusing on temporally related failures~\cite{parry2025systemic}. While repair time can also be a strong indicator of root-cause similarity, it is rarely available in practice. Benchmark datasets typically provide faulty and fixed versions of programs, but do not record the exact timestamps of repair. This becomes even more challenging in industrial datasets, where access to the full commit history is often restricted due to data privacy concerns and role-based access control~\cite{ahmedMSR}. Likewise, filtering based on repair time is infeasible in our study.

\begin{table*}[!t]
    \centering
    \resizebox{\textwidth}{!}{
    \begin{tabular}{l cccc cccc cccc cccc}
        \toprule
        \multirow{2.5}{*}{Dataset} & \multicolumn{4}{c}{$p_{\text{all}}$} & \multicolumn{4}{c}{ $p_{\text{all-preceding}}$} & \multicolumn{4}{c}{ $p_{\text{closest-by-time}}$} & \multicolumn{4}{c}{ $p_{\text{closest-time-preceding}}$}\\ 
        \cmidrule(lr){2-5} \cmidrule(lr){6-9} \cmidrule(lr){10-13} \cmidrule(lr){14-17}  
       & Min & Mean & Median & Max & Min & Mean & Median & Max & Min & Mean & Median & Max & Min & Mean & Median & Max  \\  
        \midrule
       DS1 & 656 & 656 & 656 & 656 & 0 & 328 & 328 & 656 & 66 & 66 &  66 & 66 & 0 & 62.6 & 66 & 66  \\
       DS2 & 93 & 93 & 93 & 93 & 0 & 46.5 & 46.5 & 93 & 9 & 9 & 9 & 9 & 0 & 8.5 & 9 & 9  \\
       DS3 & 27 & 27 & 27 & 27 & 0 & 13.5 & 13.5 & 27 & 3 & 3 & 3 & 3 & 0 & 2.8 & 3 & 3   \\
        \bottomrule
    \end{tabular}
    }
    \caption{Distribution of the number of test cases in corpora generated by different filtering policies for query test cases across our three datasets.}
    \label{table::searchspace_size}
\end{table*}

\begin{table*}[!t]
\centering
\resizebox{\textwidth}{!}{
\begin{tabular}{c c c
                 cccc   
                 cccc   
                 cccc   
                 cccc   
                 cccc } 
\toprule
\multirow{2.5}{*}{Dataset} & 
\multirow{2.5}{*}{Granularity} & 
\multirow{2.5}{*}{$k$} &
\multicolumn{4}{c}{Precision@$k$ (\%)} & 
\multicolumn{4}{c}{Recall@$k$ (\%)} & 
\multicolumn{4}{c}{Hit@$k$ (\%)} & 
\multicolumn{4}{c}{MAP@$k$ (\%)} & 
\multicolumn{4}{c}{MRR@$k$ (\%)} \\
\cmidrule(lr){4-7} \cmidrule(lr){8-11} \cmidrule(lr){12-15} \cmidrule(lr){16-19} \cmidrule(lr){20-23}
& & & $p_{\text{all}}$ & $p_{\text{all-pre}}$ & $p_{\text{close}}$ & $p_{\text{close-pre}}$ & $p_{\text{all}}$ & $p_{\text{all-pre}}$ & $p_{\text{close}}$ & $p_{\text{close-pre}}$ & $p_{\text{all}}$ & $p_{\text{all-pre}}$ & $p_{\text{close}}$& $p_{\text{close-pre}}$ & $p_{\text{all}}$ & $p_{\text{all-pre}}$ & $p_{\text{close}}$ & $p_{\text{close-pre}}$ & $p_{\text{all}}$ & $p_{\text{all-pre}}$ & $p_{\text{close}}$ & $p_{\text{close-pre}}$ \\
\midrule

\multirow{8}{*}{DS1}
& \multirow{4}{*}{Line}
& 1  & 54.2 & 51.6 & 54.9 & 52.1 
& 30.7 & 28.7 & 31.7 & 29.3 
& 54.2 & 51.6 & 54.9 & 52.1 
& 55.9 & 53.3 & 56.6 & 53.7 
& 55.9 & 53.3 & 56.6 & 53.7 \\
&  & 3 & 39.4 & 35.2 & 40.0 & 35.0 
& 50.0 & 43.9 & 50.7 & 45.2 
& 74.3 & 67.3 & 75.0 & 69.3 
& 65.0 & 60.5 & 65.8 & 61.6 
& 65.2 & 60.6 & 65.9 & 61.7 \\
&  & 5 & 31.3 & 27.2 & 31.4 & 28.4 
& 55.3 & 49.2 & 55.4 & 50.6 
& 77.8 & 70.8 & 77.5 & 73.4 
& 64.2 & 59.3 & 64.5 & 60.7 
& 64.6 & 59.7 & 64.7 & 60.9 \\
&  & 10 & 21.6 & 19.3 & 21.8 & 20.2 
& 62.1 & 56.6 & 64.1 & 58.7 
& 80.7 & 76.4 & 83.0 & 77.6 
& 61.8 & 58.8 & 61.9 & 59.1 
& 63.0 & 60.2 & 63.5 & 60.8 \\
\cmidrule(lr){2-23}
& \multirow{4}{*}{Unit}
& 1  & 73.7 & 70.6 & 73.4 & 70.9 
& 57.2 & 54.9 & 57.3 & 55.0 
& 73.7 & 70.6 & 73.4 & 70.9 
& 73.7 & 70.6 & 73.4 & 70.9 
& 73.7 & 70.6 & 73.4 & 70.9 \\
&  & 3 & 38.6 & 36.3 & 38.4 & 36.8 
& 74.1 & 69.5 & 73.4 & 70.5 
& 88.7 & 84.0 & 87.5 & 84.9 
& 80.0 & 76.6 & 79.4 & 77.3 
& 80.1 & 76.7 & 79.4 & 77.4 \\
&  & 5 & 26.8 & 25.0 & 26.6 & 25.4 
& 78.1 & 74.2 & 77.9 & 75.2 
& 90.4 & 87.2 & 90.1 & 88.0 
& 79.2 & 76.5 & 79.2 & 77.2 
& 79.4 & 76.6 & 79.0 & 77.2 \\
&  & 10 & 15.5 & 14.4 & 15.2 & 14.8 
& 81.8 & 78.7 & 82.2 & 79.8 
& 91.6 & 89.5 & 92.5 & 90.3 
& 77.6 & 76.8 & 77.9 & 77.2 
& 78.2 & 77.4 & 78.5 & 77.7 \\
\midrule

\multirow{8}{*}{DS2}
& \multirow{4}{*}{Line}
& 1  & 40.4 & 36.2 & 37.2 & 35.1 
& 24.6 & 21.6 & 21.2 & 19.5 
& 40.4 & 36.2 & 37.2 & 35.1 
& 40.4 & 36.2 & 37.2 & 35.1 
& 40.4 & 36.2 & 37.2 & 35.1 \\
&  & 3 & 24.1 & 19.1 & 22.7 & 21.3 
& 39.2 & 33.7 & 38.8 & 35.7 
& 57.4 & 50.0 & 58.5 & 50.0 
& 46.8 & 39.7 & 46.4 & 40.4 
& 47.5 & 40.4 & 46.6 & 41.0 \\
&  & 5 & 17.0 & 14.5 & 16.4 & 14.3 
& 45.0 & 39.0 & 41.7 & 39.3 
& 59.6 & 54.3 & 59.6 & 54.3 
& 44.9 & 38.5 & 44.2 & 37.6 
& 46.0 & 40.4 & 45.5 & 38.8 \\
&  & 10 & 9.7 & 9.3 & 10.1 & 9.4 
& 47.5 & 44.1 & 48.9 & 47.9 
& 63.8 & 59.6 & 61.7 & 58.5 
& 43.0 & 39.1 & 39.6 & 37.8 
& 45.0 & 39.4 & 41.5 & 39.5 \\
\cmidrule(lr){2-23}
& \multirow{4}{*}{Unit}
& 1  & 46.8 & 46.8 & 46.8 & 45.7 
& 35.5 & 35.2 & 35.5 & 34.1 
& 46.8 & 46.8 & 46.8 & 45.7 
& 46.8 & 46.8 & 46.8 & 45.7 
& 46.8 & 46.8 & 46.8 & 45.7 \\
&  & 3 & 26.6 & 23.8 & 26.6 & 24.1 
& 54.5 & 51.0 & 53.8 & 51.5 
& 67.0 & 62.8 & 68.1 & 62.8 
& 56.3 & 52.9 & 56.3 & 53.5 
& 56.4 & 53.4 & 56.0 & 53.7 \\
&  & 5 & 17.2 & 15.3 & 17.2 & 15.1 
& 57.3 & 52.4 & 58.2 & 52.1 
& 67.0 & 62.8 & 70.2 & 62.8 
& 54.1 & 52.3 & 56.3 & 51.4 
& 55.0 & 52.8 & 57.5 & 52.3 \\
&  & 10 & 8.3 & 9.4 & 9.5 & 8.7 
& 54.6 & 56.9 & 59.6 & 58.2 
& 69.1 & 69.1 & 72.3 & 70.2 
& 55.5 & 53.6 & 54.7 & 52.0 
& 56.2 & 54.2 & 56.0 & 53.6 \\
\midrule
\multirow{8}{*}{DS3}
& \multirow{4}{*}{Line}
& 1  & 57.1 & 53.6 & 53.6 & 46.4 
& 46.1 & 42.5 & 44.9 & 37.8 
& 57.1 & 53.6 & 53.6 & 46.4 
& 57.1 & 53.6 & 53.6 & 46.4 
& 57.1 & 53.6 & 53.6 & 46.4 \\
&  & 3 & 34.5 & 27.4 & 28.6 & 29.8 
& 66.2 & 58.0 & 64.5 & 58.8 
& 85.7 & 78.6 & 82.1 & 75.0 
& 74.4 & 66.1 & 69.3 & 63.7 
& 74.4 & 66.7 & 69.0 & 63.1 \\
&  & 5 & 22.1 & 17.9 & 20.0 & 20.0 
& 66.2 & 62.1 & 66.8 & 68.4 
& 85.7 & 78.6 & 82.1 & 82.1 
& 69.9 & 63.3 & 68.5 & 63.7 
& 69.9 & 63.3 & 68.5 & 63.2 \\
&  & 10 & 12.9 & 11.4 & 11.1 & 8.9 
& 74.5 & 69.2 & 69.8 & 59.4 
& 89.3 & 82.1 & 82.1 & 75.0 
& 68.1 & 61.9 & 67.5 & 59.3 
& 69.2 & 63.1 & 69.0 & 60.7 \\
\cmidrule(lr){2-23}
& \multirow{4}{*}{Unit}
& 1  & 82.1 & 75.0 & 75.0 & 75.0 
& 72.0 & 62.2 & 67.3 & 64.6 
& 82.1 & 75.0 & 75.0 & 75.0
& 82.1 & 75.0 & 75.0 & 75.0 
& 82.1 & 75.0 & 75.0 & 75.0 \\
&  & 3 & 44.0 & 39.3 & 36.9 & 35.7 
& 86.6 & 81.2 & 79.8 & 76.2 
& 100.0 & 92.9 & 92.9 & 89.3 
& 97.6 & 89.9 & 87.8 & 86.9 
& 98.2 & 89.3 & 87.5 & 86.9 \\
&  & 5 & 27.9 & 24.3 & 24.3 & 23.6 
& 88.1 & 80.4 & 82.4 & 79.8 
& 100.0 & 92.9 & 92.9 & 89.3 
& 94.9 & 87.3 & 87.1 & 86.0 
& 94.6 & 89.3 & 87.5 & 85.7 \\
&  & 10 & 15.7 & 13.6 & 13.2 & 11.8 
& 95.2 & 86.0 & 86.3 & 79.2 
& 100.0 & 92.9 & 92.9 & 89.3 
& 93.8 & 86.6 & 87.0 & 82.4 
& 94.6 & 87.5 & 87.5 & 83.9 \\
\bottomrule
\end{tabular}
}
\caption{The fault localization effectiveness of \ourtool at line and unit granularity levels using different filtering policies across datasets DS1, DS2, and DS3. The policies $p_{\text{all-pre}}$, $p_{\text{close}}$, and $p_{\text{close-pre}}$ correspond to $p_{\text{all-preceding}}$, $p_{\text{closest-by-time}}$, and $p_{\text{closest-time-preceding}}$, respectively.}
\label{table::filtering_accuracy}
\end{table*}

\begin{table*}[!t]
\centering
\resizebox{\linewidth}{!}{
\begin{tabular}{cc 
                cc cc cc cc      
                cc cc cc cc }
\toprule

\multirow{4.5}{*}{Dataset} & 
\multirow{4.5}{*}{$k$} &

\multicolumn{8}{c}{Average Token Count} &
\multicolumn{8}{c}{Inference Time} \\

\cmidrule(lr){3-10} 
\cmidrule(lr){11-18}

& &
\multicolumn{2}{c}{$p_{\text{all}}$} &
\multicolumn{2}{c}{$p_{\text{all-pre}}$} &
\multicolumn{2}{c}{$p_{\text{close}}$} &
\multicolumn{2}{c}{$p_{\text{close-pre}}$} &
\multicolumn{2}{c}{$p_{\text{all}}$} &
\multicolumn{2}{c}{$p_{\text{all-pre}}$} &
\multicolumn{2}{c}{$p_{\text{close}}$} &
\multicolumn{2}{c}{$p_{\text{close-pre}}$} \\

\cmidrule(lr){3-4} \cmidrule(lr){5-6} \cmidrule(lr){7-8} \cmidrule(lr){9-10} \cmidrule(lr){11-12}
\cmidrule(lr){13-14} \cmidrule(lr){15-16} \cmidrule(lr){17-18} 

& &
In & Out &
In & Out &
In & Out &
In & Out &
Avg (sec) & Sum &
Avg (sec) & Sum &
Avg (sec)& Sum &
Avg (sec) & Sum \\

\midrule

\multirow{4}{*}{DS1}
& 1  & 6.56k & 26 & 6.55k & 25 & 6.55k & 26 & 6.55k & 25 
     & 38.3 & 6h 59m 16s & 38.4 & 7h 0m 16s & 38.4 & 7h 0m 41s & 38.3 & 6h 59m 45s \\
& 3  & 6.56k & 71 & 6.55k & 70 & 6.55k & 72 & 6.55k & 71 
     & 41.6 & 7h 35m 59s & 41.6 & 7h 35m 0s & 41.7 & 7h 36m 31s & 41.6 & 7h 35m 25s \\
& 5  & 6.56k & 116 & 6.55k & 115 & 6.55k & 117 & 6.55k & 116 
     & 44.9 & 8h 11m 23s & 45.0 & 8h 12m 21s & 45.1 & 8h 14m 16s & 45.1 & 8h 13m 26s \\
& 10 & 6.56k & 224 & 6.55k & 221 & 6.56k & 222 & 6.55k & 222 
     & 52.8 & 9h 38m 37s & 52.7 & 9h 36m 36s & 52.8 & 9h 38m 25s & 52.7 & 9h 37m 36s \\

\midrule

\multirow{4}{*}{DS2}
& 1  & 9.30k & 29 & 9.30k & 30 & 9.30k & 30 & 9.30k & 30 
     & 46.5 & 1h 12m 54s & 46.6 & 1h 13m 3s & 46.8 & 1h 13m 14s & 46.5 & 1h 12m 53s \\
& 3  & 9.30k & 80 & 9.30k & 75 & 9.30k & 79 & 9.30k & 76 
     & 50.1 & 1h 18m 32s & 50.0 & 1h 18m 18s & 50.1 & 1h 18m 28s & 50.0 & 1h 18m 19s \\
& 5  & 9.30k & 130 & 9.30k & 128 & 9.30k & 125 & 9.30k & 124 
     & 53.8 & 1h 24m 21s & 53.8 & 1h 24m 21s & 53.7 & 1h 24m 7s & 53.5 & 1h 23m 48s \\
& 10 & 9.30k & 244 & 9.30k & 233 & 9.31k & 256 & 9.30k & 242 
     & 62.2 & 1h 37m 22s & 61.5 & 1h 36m 25s & 63.2 & 1h 38m 56s & 62.0 & 1h 37m 9s \\

\midrule

\multirow{4}{*}{DS3}
& 1  & 5.45k & 53 & 5.45k & 56 & 5.45k & 54 & 5.45k & 49 
     & 37.5 & 16m 15s & 37.4 & 16m 13s & 37.5 & 16m 13s & 36.9 & 15m 58s \\
& 3  & 5.45k & 194 & 5.45k & 176 & 5.45k & 152 & 5.45k & 207 
     & 47.3 & 20m 29s & 46.2 & 20m 0s & 44.5 & 19m 18s & 48.6 & 21m 4s \\
& 5  & 5.45k & 284 & 5.45k & 247 & 5.45k & 272 & 5.45k & 239 
     & 54.1 & 23m 25s & 51.1 & 22m 8s & 53.0 & 22m 58s & 50.8 & 21m 59s \\
& 10 & 5.46k & 409 & 5.45k & 396 & 5.45k & 431 & 5.45k & 387 
     & 62.6 & 27m 8s & 62.0 & 26m 52s & 64.6 & 27m 58s & 61.3 & 26m 32s \\

\bottomrule
\end{tabular}
}
\caption{The average input (In) and output (Out) token counts, along with the total and average inference time per test case, for line-level FL across datasets using different filtering policies. The policies $p_{\text{all-pre}}$, $p_{\text{close}}$, and $p_{\text{close-pre}}$ correspond to $p_{\text{all-preceding}}$, $p_{\text{closest-by-time}}$, and $p_{\text{closest-time-preceding}}$, respectively.}
\label{table::sampling_strategy_efficiency}
\end{table*}

~\autoref{table::searchspace_size} presents the distribution of the number of test cases included in the knowledge base inferred by each filtering policy for each query test case across our three datasets. To create the corpora filtered by $p_{\text{closest-by-time}}$ and $p_{\text{closest-time-preceding}}$, we retain 10\% of the test cases with the closest failure times to the query test case from the corpora generated by $p_{\text{all}}$ and $p_{\text{all-preceding}}$, respectively. As expected, the largest corpora are generated using the $p_{\text{all}}$ policy. By limiting the corpora to only include test cases that occur before the query test's failure, the $p_{\text{all-preceding}}$ policy generates corpora with an average size that is half of that produced by $p_{\text{all}}$. The corpora generated by $p_{\text{closest-by-time}}$ are 10\% the size of those generated by the no-filtering policy, while the corpora generated by $p_{\text{closest-time-preceding}}$ have a similar average size to those generated by $p_{\text{closest-by-time}}$, with minimum values of 0, which correspond to test cases whose failure is the earliest in the corpus and therefore have no temporally preceding test cases available.

Next, we evaluate the impact of using these different corpora on the effectiveness, scalability, and efficiency of FL performed by \ourtool. Specifically, the FL results demonstrate that the size and quality of the knowledge base influence the performance of the similarity search engine, thereby affecting the overall performance of \ourtool. In this experiment, all other settings, including the number of retrieved similar test cases and the annotation threshold, are set to their default values as specified in~\autoref{algorithm1}.

The FL effectiveness results for \ourtool under different filtering policies are presented in~\autoref{table::filtering_accuracy}. Overall, the no-filtering policy $p_{\text{all}}$ consistently achieves the best FL effectiveness across all three datasets. The policy $p_{\text{closest-by-time}}$ performs comparably, yielding nearly identical results on DS1, but exhibiting a modest decrease in effectiveness on DS2 and a more pronounced degradation on DS3. This trend may stem from the greater diversity of test cases in DS3, as observed in~\autoref{rq3}, which has a smaller average number of annotated lines. This increased diversity among test cases makes similarity retrieval more challenging in general, and particularly so in this setting where the knowledge base contains fewer test cases. 

Policies that exclude test cases with later failure timestamps from the knowledge base, i.e., $p_{\text{all-preceding}}$ and $p_{\text{closest-time-preceding}}$, cause a slight decrease in FL performance, highlighting the importance of including test cases in the knowledge base regardless of their temporal relation to the query test case. In general, even with the filtering policy yielding the lowest performance, \ourtool still outperforms \baseline (see~\autoref{table::accuracy::results} for a comparison with \baseline).

Because each filtering policy yields a distinct knowledge base, which may result in a different most similar test case for a given query, it can affect not only FL effectiveness but also the associated computational cost, measured in terms of token usage and inference time. 
~\autoref{table::sampling_strategy_efficiency} summarizes the results. Overall, no meaningful changes in the scalability or efficiency of \ourtool are observed across different filtering policies. This indicates that the various policies provide consistently relevant faulty test cases within the knowledge base, thereby preserving FL's effectiveness while introducing only minor variations in the computational efficiency of FL.

Given that temporal filtering policies reduce the knowledge base size with minimal compromise to retrieval quality, leveraging correlations among test cases that fail around the same time is an effective approach when knowledge base shrinkage is desirable. In such cases, based on our results, $p_{\text{closest-by-time}}$ is generally preferred over $p_{\text{closest-time-preceding}}$, as it applies less aggressive pruning to the corpus of fault-labeled test cases, potentially preserving more relevant candidates.

\Finding{\ourtool outperforms \baseline regardless of the filtering policy used to construct the knowledge base. Including test cases in the knowledge base, regardless of their temporal relationship to the query test case, further improves \ourtool's effectiveness. Therefore, both $p_{\text{all}}$ and $p_{\text{closest-by-time}}$ are suitable filtering policies, with $p_{\text{closest-by-time}}$ achieving nearly identical performance while reducing the knowledge base size by almost 90\%.}

\section{Discussion}\label{sec:discussion}
\begin{table}[!t]
    \centering
    \scriptsize
    \resizebox{0.98\columnwidth}{!}{
    \begin{tabular}{c c c c c c}
        \toprule
         TCFL Mode & Filtering & Similarity Search & Retrieval & Annotation & Prompt Design\\
        \midrule
        Default & False & True & True & True & Baseline \\
        \midrule
        Random & False & False & True & True & Baseline \\
        \midrule
        Annotation-Free & False & True & True & False & $\,\,\,$Baseline$^+$ \\
        \midrule
        Directive & False & True & True & True & Directive \\
        \bottomrule
    \end{tabular}
    }
    \caption{Different \ourtool's modes used to investigate the impact of design choices. Baseline, Baseline$^+$, and Directive prompt templates are presented in~\autoref{fig:prmpt_template},~\autoref{fig:prmpt_no_annotation}, and~\autoref{fig:directive:template}, respectively.}
    \label{table::ablation::design}
\end{table}

In this section, we further investigate the effects of key design choices on the performance of \ourtool. We also discuss additional insights and potential directions for future work.

\subsection{Ablation Study: Evaluating Key Design Choices}\label{sub::ablation}
To assess the impact of key design choices, we define four distinct FL modes, as shown in~\autoref{table::ablation::design}, and conduct experiments for each to compare their results. These FL modes are designed to isolate and evaluate the impact of a single component of \ourtool at a time and are defined as follows:
\begin{enumerate}
    \item Default: This mode strictly follows ~\autoref{algorithm1}, applying no filtering by using the $p_{\text{all}}$ policy. It performs similarity search as described in ~\autoref{sub::sim::engine} to identify the most similar example, includes an annotator module with a threshold of 0.05, and uses the \baseline's prompt template shown in ~\autoref{fig:prmpt_template}. This configuration served as the primary setting against which we compared \baseline when addressing RQ1 and RQ2, with results presented in ~\autoref{table::accuracy::results} and ~\autoref{table::efficiency::results}.
    \item Random: In this mode, the similarity search is disabled, and a single test case is retrieved randomly from the knowledge base without considering similarity. All other settings remain the same as in the Default mode. This configuration allows us to evaluate the impact of similarity-based retrieval on \ourtool's performance.
    \item Annotation-Free: In this mode, the annotator module is disabled, while all other settings remain the same as in the Default mode. Here, query test lines that are similar to the retrieved faulty lines are not annotated. Instead, the retrieved faulty lines are presented as a separate section in the prompt, with instructions referring to them, as illustrated in~\autoref{fig:prmpt_no_annotation}. This configuration allows us to assess the impact of the annotator module on \ourtool's performance.
    \item Directive: In this mode, all components remain the same as in the Default mode, except that \ourtool's prompt template (the template recommended by \baseline) is replaced with the template shown in~\autoref{fig:directive:template}, which we refer to as the directive prompt template. This template explicitly references the annotator module's outcomes, namely, the annotation message (i.e., \texttt{`\# !!! high likelihood of being faulty !!!'}) and instructs the LLM to begin its analysis by examining the lines marked with the message. Furthermore, it reverses the order of the Task Instructions and Inputs sections to emphasize the importance of the instructions, including the annotation guidance. This configuration allows us to explore how an alternative prompt design affects \ourtool's performance through prompt engineering.
\end{enumerate}

In addition to these modes, another design choice is the filtering policy, which we investigated through multiple variations, including the absence of a filtering module ($p_{\text{all}}$), as discussed in RQ4 in~\autoref{subsub::rq4}.

\begin{figure}[!t]
  \centering
  \scriptsize
  \framebox[\columnwidth][l]{\parbox{0.95\columnwidth}{
  \Large \textbf{Task Description} \vspace{0.4em} \\
  \normalsize As an expert software engineer and tester, your mission is to localize faults in \{\emph{programming\_language\}} test scripts at the \emph{\{element\}} level. You will be provided with the test scripts and the error message caused by the test failure. Your goal is to identify \emph{\{k\}} \emph{\{element\}}s that are most likely responsible for the failure and require modification. \textbf{\hl{To reason about this faulty test case, you will also be provided with a set of faulty lines retrieved from similar test scripts.}}\vspace{0.6em}\\
 \Large \textbf{Inputs} \vspace{0.4em} \\
\large \textbf{Error Message}
 \vspace{0.4em}\\
\normalsize Here is the error message caused by the test failure:\vspace{0.4em}\\
  \emph{\{err\_msg\}}\vspace{0.4em}\\
\large \textbf{Code}\vspace{0.4em}\\
\normalsize Below are the \emph{\{programming\_language\}} test scripts:\vspace{0.4em}\\
   \emph{\{test\_code\}}\vspace{0.6em}\\
\large \textbf{\hl{Additional Context}}
 \vspace{0.4em}\\
 \normalsize \textbf{\hl{Below is a set of faulty lines that caused a similar error message in a similar faulty \emph{\{programming\_language\}} test case:}}\vspace{0.4em}\\
 \emph{\textbf{\hl{\{similar\_faulty\_lines\}}}}\vspace{0.6em}\\
\Large\textbf{Task Instructions}\vspace{0.4em}
\normalsize
\begin{enumerate}[leftmargin=15pt]
\item Carefully examine the provided test scripts and the associated error message \textbf{\hl{and the similar faulty lines provided as additional context}}.
    \item Identify the \emph{\{k\}} \emph{\{element\}}s that are most likely to contain the faults.
    \item Return a list of faulty \emph{\{element\}}s and their \emph{\{ID\}}s, without any additional explanation. Note that the list of \emph{\{element\}}s and their \emph{\{ID\}}s should be within the range 1 to \emph{\{max\_element\_id\}} and the size of the list must be exactly \emph{\{k\}}. The list should be also in descending order of likelihood of containing the fault, with the most suspicious \emph{\{element\}} first and the least suspicious \emph{\{element\}} last. Ensure that your response is strictly in the specified format. The output should follow this format: \emph{\{output\_template\}}
\end{enumerate}
  }}
  \caption{Prompt template for \ourtool with the annotator module disabled, in which faulty lines retrieved from similar test cases are included as a separate section. The highlighted text presents additional instructions and information included in this template that are not present in the \baseline's prompt template shown in~\autoref{fig:prmpt_template}.}
  \label{fig:prmpt_no_annotation}
\end{figure}
  
\begin{figure}[!t]
  \centering
  \scriptsize
  \framebox[\columnwidth][l]{\parbox{0.95\columnwidth}{
  \Large \textbf{Task Description} \vspace{0.4em} \\
  \normalsize As an expert software engineer and tester, your mission is to localize faults in \{\emph{programming\_language\}} test scripts at the \emph{\{element\}} level. You will be provided with the test scripts and the error message caused by the test failure. Your goal is to identify \emph{\{k\}} \emph{\{element\}}s that are most likely responsible for the failure and require modification.\vspace{0.6em}\\
  \Large\textbf{Task Instructions}\vspace{0.4em}
\normalsize
\begin{enumerate}[leftmargin=15pt]
\item Identify \{k\} \{element\}s in the following test script that are likely to contain the fault.
\item You must pay attention to the lines marked with \textbf{\textit{`\# !!! high likelihood of being faulty !!!'}} and start with investigating them first.
  \item Return a list of faulty \emph{\{element\}}s and their \emph{\{ID\}}s, without any additional explanation. Note that the list of \emph{\{element\}}s and their \emph{\{ID\}}s should be within the range 1 to \emph{\{max\_element\_id\}} and the size of the list must be exactly \emph{\{k\}}. The list should be also in descending order of likelihood of containing the fault, with the most suspicious \emph{\{element\}} first and the least suspicious \emph{\{element\}} last. Ensure that your response is strictly in the specified format. The output should follow this format: \emph{\{output\_template\}}
\end{enumerate}
 \Large \textbf{Inputs} \vspace{0.4em} \\
\large \textbf{Error Message}
 \vspace{0.4em}\\
\normalsize Here is the error message caused by the test failure:\vspace{0.4em}\\
  \emph{\{err\_msg\}}\vspace{0.4em}\\
\large \textbf{Code}\vspace{0.4em}\\
\normalsize Below are the \emph{\{programming\_language\}} test scripts:\vspace{0.4em}\\
   \emph{\{test\_code\}}\\
  }}
  \caption{A directive prompt template for test code fault localization, explicitly instructing the LLM to start with investigating the annotated lines. }
  \label{fig:directive:template}
\end{figure}

The FL effectiveness results for the four FL modes described above are presented in~\autoref{table::ablation_effectiveness_results}.

\subsubsection{Effectiveness of different \ourtool's modes}

\noindent\textbf{Default vs. Random Modes.} Across all datasets and granularity levels, the Default mode consistently outperforms the Random mode across all metrics, highlighting the importance of similarity-based retrieval in \ourtool. For instance, at the line level for test cases in DS1, Precision@1 increases from 44.4\% in Random mode to 54.2\% in Default mode. Precision@1 for line-level FL on DS2 rises to almost twice the Random mode value, from 21.3\% to 40.4\% in the Default mode. A similar pattern appears across other metrics; for instance, Recall@5 improves from 21.6\% in Random to 45.0\% in Default. Similarly, at the unit level, Default mode achieves higher performance, with Precision@1 increasing from 67.0\% to 73.7\% on DS1 and Precision@3 from 28.6\% to 44.0\% on DS3. The performance gap is generally larger at line-level granularity, indicating that similarity-based test case retrieval is especially beneficial for finer-grained FL. The results confirm that the similarity search component is a key driver of \ourtool's effectiveness, as random retrieval significantly lags behind the Default configuration and performs comparably to the \baseline approach.

\noindent\textbf{Default vs. Annotation-Free Modes.} The comparison between the Default and Annotation-Free modes also reveals differences in TCFL effectiveness. For example, at the line level in DS2, Precision@1 improves from 28.7\% in the Annotation-Free mode to 40.4\%. On DS3, line-level Precision@3 rises from 23.8\% in Annotation-Free mode to 34.5\% in Default mode.
The performance gap between the two modes persists across other metrics and at the unit level as well. These results suggest that excluding the annotator module reduces TCFL's effectiveness across all datasets and granularity levels, underscoring the importance of annotating retrieved faulty lines rather than directly providing information from retrieved similar test cases as input to the LLM. Furthermore, a comparison between the Annotation-Free and Random modes indicates that they have a comparable impact on \ourtool's performance.

\begin{table*}[!t]
\centering
\resizebox{\textwidth}{!}{
\begin{tabular}{c c c
                 cccc   
                 cccc   
                 cccc   
                 cccc   
                 cccc } 
\toprule
\multirow{2.5}{*}{Dataset} & 
\multirow{2.5}{*}{Granularity} & 
\multirow{2.5}{*}{$k$} &
\multicolumn{4}{c}{Precision@$k$ (\%)} & 
\multicolumn{4}{c}{Recall@$k$ (\%)} & 
\multicolumn{4}{c}{Hit@$k$ (\%)} & 
\multicolumn{4}{c}{MAP@$k$ (\%)} & 
\multicolumn{4}{c}{MRR@$k$ (\%)} \\
\cmidrule(lr){4-7} \cmidrule(lr){8-11} \cmidrule(lr){12-15} \cmidrule(lr){16-19} \cmidrule(lr){20-23}
& & & Def & Rand & Ann-Free & Direct & Def & Rand & Ann-Free & Direct & Def & Rand & Ann-Free & Direct & Def & Rand & Ann-Free & Direct & Def & Rand & Ann-Free & Direct \\
\midrule

\multirow{8}{*}{DS1}

& \multirow{4}{*}{Line}
& 1  & 54.2 & 44.4 & 46.1 & 49.5 &
      30.7 & 24.2 & 25.2 & 27.4 &
      54.2 & 44.4 & 46.1 & 49.5 &
      55.9 & 46.1 & 47.8 & 51.1 &
      55.9 & 46.1 & 47.8 & 51.1 \\
& & 3  & 39.4 & 27.2 & 30.7 & 40.0 &
      50.0 & 34.0 & 38.4 & 50.4 &
      74.3 & 57.4 & 63.0 & 75.6 &
      65.0 & 51.7 & 54.9 & 65.4 &
      65.2 & 51.9 & 54.9 & 65.3 \\
& & 5  & 31.3 & 21.3 & 24.4 & 31.2 &
      55.3 & 39.1 & 43.9 & 55.7 &
      77.8 & 62.4 & 68.2 & 78.1 &
      64.2 & 52.2 & 54.9 & 64.5 &
      64.6 & 52.7 & 55.2 & 64.7 \\
& & 10 & 21.6 & 15.9 & 18.0 & 21.6 &
      62.1 & 46.7 & 52.3 & 61.5 &
      80.7 & 68.2 & 73.4 & 79.8 &
      61.8 & 51.9 & 53.8 & 62.5 &
      63.0 & 53.9 & 55.3 & 63.6 \\
\cmidrule{2-23}
& \multirow{4}{*}{Unit}
& 1  & 73.7 & 67.0 & 68.2 & 70.0 &
      57.2 & 51.9 & 52.6 & 54.0 &
      73.7 & 67.0 & 68.2 & 70.0 &
      73.7 & 67.0 & 68.2 & 70.0 &
      73.7 & 67.0 & 68.2 & 70.0 \\
& & 3  & 38.6 & 32.4 & 35.2 & 38.3 &
      74.1 & 64.0 & 68.2 & 73.6 &
      88.7 & 79.5 & 83.1 & 88.6 &
      80.0 & 72.5 & 75.0 & 80.0 &
      80.1 & 72.5 & 75.0 & 80.0 \\
& & 5  & 26.8 & 22.0 & 24.2 & 26.5 &
      78.1 & 67.9 & 72.5 & 78.0 &
      90.4 & 81.7 & 86.8 & 91.2 &
      79.2 & 72.9 & 75.9 & 80.6 &
      79.4 & 73.1 & 75.9 & 80.8 \\
& & 10 & 15.5 & 13.3 & 14.3 & 14.9 &
      81.8 & 74.2 & 78.2 & 81.1 &
      91.6 & 86.1 & 90.0 & 91.8 &
      77.6 & 73.9 & 75.5 & 79.3 &
      78.2 & 74.4 & 76.2 & 79.8 \\
\midrule
\multirow{8}{*}{DS2}

& \multirow{4}{*}{Line}
& 1  & 40.4 & 21.3 & 28.7 & 45.7 &
      24.6 & 12.7 & 18.8 & 28.3 &
      40.4 & 21.3 & 28.7 & 45.7 &
      40.4 & 21.3 & 28.7 & 45.7 &
      40.4 & 21.3 & 28.7 & 45.7 \\
& & 3  & 24.1 & 13.8 & 18.4 & 27.7 &
      39.2 & 21.6 & 32.3 & 47.9 &
      57.4 & 33.0 & 45.7 & 62.8 &
      46.8 & 26.9 & 35.2 & 53.7 &
      47.5 & 26.6 & 35.1 & 54.3 \\
& & 5  & 17.0 & 8.1 & 11.1 & 18.1 &
      45.0 & 21.6 & 30.3 & 51.9 &
      59.6 & 34.0 & 44.7 & 66.0 &
      44.9 & 25.6 & 31.0 & 52.6 &
      46.0 & 25.5 & 31.4 & 54.0 \\
& & 10 & 9.7 & 7.0 & 7.2 & 10.3 &
      47.5 & 31.6 & 36.9 & 55.8 &
      63.8 & 44.7 & 51.1 & 69.1 &
      43.0 & 28.9 & 29.7 & 52.6 &
      45.0 & 29.2 & 30.2 & 53.7 \\
\cmidrule{2-23}
& \multirow{4}{*}{Unit}
& 1  & 46.8 & 44.7 & 44.7 & 52.1 &
      35.5 & 33.6 & 33.6 & 39.1 &
      46.8 & 44.7 & 44.7 & 52.1 &
      46.8 & 44.7 & 44.7 & 52.1 &
      46.8 & 44.7 & 44.7 & 52.1 \\
& & 3  & 26.6 & 20.9 & 22.3 & 27.7 &
      54.5 & 42.9 & 48.5 & 57.5 &
      67.0 & 55.3 & 61.7 & 68.1 &
      56.3 & 49.6 & 51.8 & 59.2 &
      56.4 & 49.5 & 52.3 & 59.4 \\
& & 5  & 17.2 & 12.6 & 14.3 & 17.4 &
      57.3 & 42.2 & 47.8 & 60.8 &
      67.0 & 56.4 & 61.7 & 70.2 &
      54.1 & 48.4 & 50.7 & 58.1 &
      55.0 & 48.2 & 52.1 & 59.3 \\
& & 10 & 8.3 & 7.4 & 7.8 & 9.3 &
      54.6 & 47.3 & 51.5 & 63.0 &
      69.1 & 60.6 & 63.8 & 74.5 &
      55.5 & 49.5 & 50.3 & 58.2 &
      56.2 & 50.5 & 52.1 & 59.2 \\
\midrule
\multirow{8}{*}{DS3}

& \multirow{4}{*}{Line}
& 1  & 57.1 & 42.9 & 42.9 & 60.7 &
      46.1 & 34.2 & 34.2 & 47.9 &
      57.1 & 42.9 & 42.9 & 60.7 &
      57.1 & 42.9 & 42.9 & 60.7 &
      57.1 & 42.9 & 42.9 & 60.7 \\
& & 3  & 34.5 & 23.8 & 23.8 & 36.9 &
      66.2 & 49.0 & 56.1 & 71.6 &
      85.7 & 60.7 & 64.3 & 89.3 &
      74.4 & 50.9 & 52.4 & 74.4 &
      74.4 & 50.6 & 52.4 & 75.0 \\
& & 5  & 22.1 & 13.6 & 14.3 & 22.9 &
      66.2 & 51.1 & 54.7 & 72.2 &
      85.7 & 60.7 & 71.4 & 89.3 &
      69.9 & 50.8 & 53.7 & 68.1 &
      69.9 & 50.3 & 53.7 & 69.9 \\
& & 10 & 12.9 & 8.6 & 8.9 & 13.9 &
      74.5 & 60.6 & 64.8 & 75.6 &
      89.3 & 71.4 & 82.1 & 85.7 &
      68.1 & 52.2 & 53.5 & 65.1 &
      69.2 & 52.8 & 55.3 & 64.7 \\
\cmidrule{2-23}
& \multirow{4}{*}{Unit}
& 1  & 82.1 & 71.4 & 71.4 & 82.1 &
      72.0 & 63.7 & 63.7 & 70.2 &
      82.1 & 71.4 & 71.4 & 82.1 &
      82.1 & 71.4 & 71.4 & 82.1 &
      82.1 & 71.4 & 71.4 & 82.1 \\
& & 3  & 44.0 & 28.6 & 27.4 & 45.2 &
      86.6 & 68.2 & 72.6 & 88.7 &
      100.0 & 78.6 & 82.1 & 100.0 &
      97.6 & 76.8 & 75.0 & 96.1 &
      98.2 & 76.8 & 75.0 & 95.8 \\
& & 5  & 27.9 & 17.1 & 18.6 & 30.0 &
      88.1 & 67.3 & 74.4 & 92.3 &
      100.0 & 75.0 & 89.3 & 100.0 &
      94.9 & 73.9 & 77.0 & 91.8 &
      94.6 & 73.2 & 76.9 & 91.1 \\
& & 10 & 15.7 & 9.6 & 10.0 & 14.6 &
      95.2 & 71.1 & 78.0 & 90.5 &
       100.0 & 85.7 & 92.9 & 96.4 &
      93.8 & 77.1 & 78.5 & 90.1 &
      94.6 & 76.5 & 78.4 & 89.3 \\
      \bottomrule
\end{tabular}
}
\caption{The fault localization effectiveness of \ourtool at line and unit granularity levels using the FL modes defined in~\autoref{table::ablation::design} across datasets DS1, DS2, and DS3. \emph{Def}, \emph{Rand}, \emph{Ann-Free}, and \emph{Direct} correspond to the default, random, annotation-free, and directive modes, respectively.}
\label{table::ablation_effectiveness_results}
\end{table*}

\begin{table*}[!t]
\centering
\resizebox{\linewidth}{!}{
\begin{tabular}{cc 
                cc cc cc cc      
                cc cc cc cc }
\toprule

\multirow{4.5}{*}{Dataset} & 
\multirow{4.5}{*}{$k$} &

\multicolumn{8}{c}{Average Token Count} &
\multicolumn{8}{c}{Inference Time} \\

\cmidrule(lr){3-10} 
\cmidrule(lr){11-18}

& &
\multicolumn{2}{c}{Def} &
\multicolumn{2}{c}{Rand} &
\multicolumn{2}{c}{Ann-Free} &
\multicolumn{2}{c}{Direct} &
\multicolumn{2}{c}{Def} &
\multicolumn{2}{c}{Rand} &
\multicolumn{2}{c}{Ann-Free} &
\multicolumn{2}{c}{Direct} \\

\cmidrule(lr){3-4} \cmidrule(lr){5-6} \cmidrule(lr){7-8} \cmidrule(lr){9-10} \cmidrule(lr){11-12}
\cmidrule(lr){13-14} \cmidrule(lr){15-16} \cmidrule(lr){17-18} 

& &
In & Out &
In & Out &
In & Out &
In & Out &
Avg (sec) & Sum &
Avg (sec) & Sum &
Avg (sec)& Sum &
Avg (sec) & Sum \\

\midrule

\multirow{4}{*}{DS1}
& 1  & 6.56k & 26 & 6.51k & 24 & 6.65k & 24 & 6.57k & 27 & 38.3 & 6h 59m 16s & 38.1 & 6h 57m 21s & 38.5 & 7h 1m 6s & 38.5 & 7h 1m 25s \\
& 3  & 6.56k & 71 & 6.51k & 68 & 6.65k & 70 & 6.57k & 76 & 41.6 & 7h 35m 59s & 41.3 & 7h 32m 41s & 41.8 & 7h 37m 13s & 42.0 & 7h 39m 44s \\
& 5  & 6.56k & 116 & 6.51k & 111 & 6.65k & 113 & 6.57k & 121 & 44.9 & 8h 11m 23s & 44.6 & 8h 8m 38s & 45.0 & 8h 13m 4s & 45.5 & 8h 18m 35s \\
& 10 & 6.56k & 224 & 6.51k & 217 & 6.66k & 217 & 6.58k & 225 & 52.8 & 9h 38m 37s & 52.5 & 9h 34m 53s & 52.7 & 9h 36m 47s & 53.0 & 9h 40m 2s \\

\midrule

\multirow{4}{*}{DS2}
& 1  & 9.30k & 29 & 9.28k & 30 & 9.41k & 28 & 9.31k & 32 & 46.5 & 1h 12m 54s & 46.6 & 1h 12m 59s & 46.7 & 1h 13m 10s & 46.7 & 1h 13m 11s \\
& 3  & 9.30k & 80 & 9.28k & 80 & 9.41k & 77 & 9.31k & 82 & 50.1 & 1h 18m 32s & 50.1 & 1h 18m 32s & 50.3 & 1h 18m 43s & 50.4 & 1h 18m 59s \\
& 5  & 9.30k & 130 & 9.28k & 132 & 9.41k & 160 & 9.31k & 134 & 53.8 & 1h 24m 21s & 54.1 & 1h 24m 41s & 56.4 & 1h 28m 23s & 54.0 & 1h 24m 40s \\
& 10 & 9.30k & 244 & 9.28k & 245 & 9.41k & 228 & 9.32k & 242 & 62.2 & 1h 37m 22s & 62.2 & 1h 37m 23s & 61.3 & 1h 36m 0s & 62.2 & 1h 37m 26s \\

\midrule

\multirow{4}{*}{DS3}
& 1  & 5.45k & 53 & 5.44k & 65 & 5.58k & 71 & 5.47k & 61 & 37.5 & 16m 15s & 38.1 & 16m 30s & 39.1 & 16m 56s & 37.8 & 16m 22s \\
& 3  & 5.45k & 194 & 5.44k & 165 & 5.58k & 171 & 5.47k & 195 & 47.3 & 20m 29s & 45.4 & 19m 39s & 46.1 & 19m 58s & 47.6 & 20m 38s \\
& 5  & 5.45k & 284 & 5.44k & 276 & 5.58k & 250 & 5.47k & 273 & 54.1 & 23m 25s & 53.4 & 23m 9s & 51.7 & 22m 24s & 53.1 & 22m 59s \\
& 10 & 5.46k & 409 & 5.45k & 403 & 5.58k & 401 & 5.47k & 417 & 62.6 & 27m 8s & 62.2 & 26m 57s & 62.5 & 27m 4s & 63.7 & 27m 37s \\

\bottomrule

\end{tabular}
}
\caption{The average input (In) and output (Out) token counts, along with the total and average inference time per test case, for line-level FL across datasets using different FL modes defined in~\autoref{table::ablation::design} across datasets DS1, DS2, and DS3. \emph{Def}, \emph{Rand}, \emph{Ann-Free}, and \emph{Direct} correspond to the default, random, annotation-free, and directive modes, respectively.}
\label{table::ablation::efficiency::results}
\end{table*}

\noindent\textbf{Default vs. Directive Modes.} Comparing the effectiveness of the Default and Directive modes reveals that variations in prompt templates do not dramatically alter the results when the prompt is guided by annotations derived from retrieved relevant patterns, i.e., faulty lines. In fact, the Directive mode performs closely to the Default mode, demonstrating the benefits of retrieved annotations and highlighting \ourtool's robustness when provided with appropriately recommended structured inputs, instructions, and contextual guidance. However, the impact is dataset-dependent to some extent. In DS1, the Directive mode performs similarly to, or slightly worse than, the Default mode overall. In contrast, on DS2 and DS3, the Directive mode performs comparably to, or slightly better than, the Default mode. For example, in DS1 at \(k=1\), Directive achieves a line-level precision of 49.5\%, compared to Default's 54.2\%.
On DS2, the Directive mode achieves a line-level Precision@1 of 45.7\% compared to 40.4\% for the Default mode.

\subsection{Computational efficiency of different SPARK's modes}
~\autoref{table::ablation::efficiency::results} presents average token count and inference time of \ourtool across four various introduced modes.
In terms of scalability, the Annotation-Free mode has the highest average number of input tokens. This is expected, as it allocates a dedicated prompt section to include the content of similar faulty lines rather than relying on a lightweight annotation message for existing query test lines (see~\autoref{fig:prmpt_no_annotation}). Consequently, its higher token usage and lower effectiveness highlight the importance of the annotator module in \ourtool. The Directive template also uses slightly more tokens than Default due to its more detailed instructions (step 2 in~\autoref{fig:directive:template}). This modest increase, together with the comparable effectiveness to the Default mode, demonstrates the robustness of \ourtool in properly guiding the LLM with different prompt templates annotated with retrieved patterns. The Random mode yields the fewest input tokens. This reflects the fact that fewer query test code lines are annotated in Random mode than in Default mode, since the randomly selected test case is often not sufficiently similar to the query test, and the annotator module therefore identifies only a limited number of relevant lines. However, this apparent reduction in token usage comes at the cost of reduced effectiveness, with performance approaching that of the \baseline.
The average number of output tokens and inference time are similar across all modes, with no significant differences.

\subsection{When \ourtool Works Well and When It Does Not}\label{sub::lessons}

Our experiments provide several insights into the conditions under which \ourtool is most effective, as well as scenarios where its benefits are less pronounced. Overall, \ourtool consistently improves FL effectiveness across all datasets and evaluation metrics compared to \baseline.

We observe that effectiveness gains are more pronounced for larger and more complex test cases. In datasets such as DS2, which contain longer test scripts and a more extensive FL search space, \ourtool yields substantial improvements. This suggests that annotating relevant contextual patterns is particularly beneficial when the LLM must reason over a large number of candidate lines.

Our technique requires access to accumulated debugging knowledge, which is typically available in practical systems. However, the quality and quantity of debugging cases may depend on the type of system. For example, in relatively new systems, such accumulated debugging knowledge may be limited, which can reduce the benefits of our technique. Nevertheless, even when no such knowledge is available, our technique performs at least as well as the baseline. 

The temporal feature of test cases in the fault-labeled knowledge base also plays a role in determining performance. Studying different filtering policies shows that policies retaining both sets of test cases with failure times before and after the query test's failure generally yield better results. While bug triage~\cite{Hu2014bugtriage}, which is commonly used in practice, enables access to both sets, it may not be adopted in relatively small systems. Nevertheless, despite slightly better results, the differences in overall FL performance across policies are negligible, suggesting that similarity between test cases plays a more significant role than temporal proximity within the corpus. Therefore, regardless of the filtering policy, our technique still outperforms the baseline, demonstrating the benefit of leveraging debugging knowledge even when it is scarce or has limited temporal coverage. 

Finally, we have demonstrated that constructing knowledge bases and using their context in FL incurs negligible overhead. However, in real-world CI settings, the decision to maintain and leverage such corpora may ultimately depend on cost-benefit trade-offs and warrants further investigation in future work.


\subsection{Future Directions}\label{sub::future}
Several directions for future work can be explored. 
First, we currently retrieve a single similar test case from the knowledge base for each query test case. While effective, retrieving multiple similar test cases and extracting relevant patterns from all of them could reveal the impact of the number of retrieved examples on \ourtool's performance.

Second, we construct knowledge bases using failure timestamps; however, other factors, such as repair times when available, could be used to generate alternative knowledge bases and to study their effects on similarity search. Furthermore, alternative similarity retrieval strategies beyond the current one, i.e., cosine similarity with approximate k-nearest neighbors, may further enhance performance.

Third, one can explore different levels of annotation granularity as another avenue of research. At present, we measure the distance between the complete query test code lines and the retrieved faulty lines to annotate the entire lines. This could be extended to coarser levels, such as CFG blocks, or finer levels, such as individual tokens.

Fourth, we currently employ a lightweight textual similarity based on normalized Levenshtein distance to identify near-identical assertion patterns, API usage constructs, and recurring test logic, while relying on the LLM for semantic reasoning. Other distance metrics, such as embeddings of grammar rule types or code semantics, can be used to capture syntactic or semantic similarity and yield more informative relevance scores.

Fifth, extending the retriever module's outputs beyond faulty lines to include additional data modalities retrieved from similar test cases, such as code diffs, error messages, or relevant test code fragments, represents another direction for future research. While such extensions could enrich the contextual information provided to the LLM, they would require new strategies for extracting and annotating relevant patterns within the query test for effective TCFL.

Finally, we evaluated \ourtool on diverse datasets of Python test cases with varying sizes. Extending the evaluation to test cases written in other programming languages is a natural direction for future research.

\section{Threats to validity}\label{sec:threats}
In this section, we identify and discuss potential threats to the validity of our study, focusing on factors that may affect the internal, external, and construct validity of our results.

\subsection{Internal Validity}
As discussed in~\autoref{sec:method}, accumulated debugging knowledge may not always exhibit strict temporal dependency in practice. A potential threat arises from leveraging failure timestamps to filter the knowledge base, as failure time does not always correlate with root-cause similarity. To mitigate this threat, we evaluate multiple filtering policies, including settings that simulate using all available fault-labeled test cases and temporally constrained subsets.

Although faults can be localized regardless of when they fail (due to the bug triage process), using failures that occur after the query test case may still introduce information leakage, since their labels might not be available in some real debugging scenarios. To address this, we implement preceding-only policies, namely $p_{\text{all-preceding}}$ and $p_{\text{closest-time-preceding}}$, which restrict retrieval to historically available failures.  
This design enables evaluation under different levels of debugging knowledge availability, covering both scenarios where future labeled failures are accessible and scenarios where only historically available failures can be used.

The effectiveness and robustness of \ourtool may be influenced by configuration choices, including the annotation distance threshold, the filtering policy, and prompt construction. To mitigate this internal validity threat, we conduct sensitivity analyses across a range of annotation thresholds at varying levels of strictness, evaluate multiple filtering policies, and perform ablation studies to isolate the contributions of individual system components.

\subsection{External Validity}
A potential threat to the generalizability of \ourtool is that performance may vary across software systems, programming languages, and testing environments. To mitigate this concern, we evaluated \ourtool on three diverse system-level test script datasets, each intended to test a different SUT. These datasets, comprising 657, 94, and 28 system-level test cases, respectively, are actively used by our industry partner, thereby ensuring realistic, practically relevant evaluation scenarios.

Regarding external validity, a potential threat arises from data contamination if the study subjects were included in the LLM's training corpus. We mitigate this risk by using proprietary industrial datasets that are not publicly accessible, thereby minimizing the likelihood that the model has prior exposure to the specific faults or system architectures under evaluation.

\subsection{Construct Validity}
Identifying the true faulty locations in FL studies is challenging because a single commit may contain multiple changes, some of which are not relevant to the observed failure (e.g., refactoring changes). To mitigate this threat, we adopt the semi-automated labeling approach of Saboor et al.~\cite{saboor2025black}, described in~\autoref{sub::bench}, which combines outlier exclusion with developer-verified refinement to reduce false positive fault locations.

As an LLM-based approach, \ourtool's effectiveness may depend on the reasoning and instruction-following capabilities of the underlying language model. To reduce this threat, we employ the same recommended model as \baseline, with identical hyperparameter settings (e.g., temperature), ensuring a fair and controlled comparison between the two approaches.

A further threat concerns the distribution of similarity among faulty test cases in the debugging knowledge corpus. If many test cases share overlapping statements or structural similarities, retrieving the most similar faulty-labeled test case may become less discriminating and approach random selection. To mitigate this threat, we evaluate \ourtool across three datasets that differ in the degree of similarity between the query test case and the retrieved faulty cases, as shown in~\autoref{table::annotation::count}. In addition, we explore the impact of the similarity search engine module by comparing it with a random-selection strategy from the debugging knowledge corpus.

\section{Related Work}\label{sec:rw}
In this section, we review prior research studies on FL, tracing its evolution from traditional techniques to recent advances leveraging LLMs, as well as approaches that enhance LLMs through domain-specific contextual information.

Traditional FL methods are predominantly based on program spectra, where statistical correlations between test coverage and observed failures are used to localize faults. These methods, commonly referred to as Spectrum-Based Fault Localization (SBFL)~\cite{DBLP:journals/ieicetd/ZhengHCYFX24, DBLP:journals/jss/RaselimoF24, DBLP:journals/access/SarhanB22,de2016spectrum,zakari2020spectrum}, assign suspiciousness scores to program elements (e.g., statements or blocks) based on their execution frequency in failing versus passing test cases. While simple and computationally efficient, SBFL is limited in its ability to capture complex semantic relationships between code and faults.

To address these limitations, machine learning (ML) techniques have been applied to FL, enabling models to learn patterns from program execution and associated artifacts. Early approaches relied on textual similarity, measuring overlaps between bug reports and source code artifacts~\cite{text_retrieval,traceability2018}. 
Subsequent deep learning approaches capture richer semantic relationships. For instance, Huo et al. proposed NP-CNN~\cite{NP-CNN} (Natural language and Programming language Convolutional Neural Network), which extracts both lexical and structural features from bug reports and source code through convolutional layers, and fuses them into a unified representation to more accurately identify the source files associated with a given bug report.
Wang et al. introduced MD-CNN~\cite{MD-CNN}, a Multi-Dimension Convolutional Neural Network for FL. This model extracts five statistical feature dimensions from bug reports and source files and leverages a CNN to capture complex nonlinear correlations among them. Yang et al. developed MRAM~\cite{MRAM}, a hybrid model combining recurrent neural networks (RNNs) with attention mechanisms, integrating method-structured features such as token sequences, API calls, and comments with bug-fixing features from code revision graphs to enhance method-level FL. Lam et al. proposed DNNLOC~\cite{lam17}, a hybrid approach combining a revised Vector Space Model (rVSM) with a deep neural network (DNN) for file-level fault localization. rVSM extracts textual similarity features between bug reports and source files, while the DNN bridges lexical gaps by relating abstract terms in bug reports to semantically relevant tokens in the code. More recently, Zhang et al. introduced BugRadar~\cite{zhang23}, which integrates a knowledge graph with hyperbolic attention embeddings, enriched textual similarity features, and history-based data derived from past bug reports and fixing-time information to improve FL.

Beyond textual similarity, some studies leverage program execution and structural information. GRACE~\cite{DBLP:conf/sigsoft/LouZDLSHZZ21} models code coverage as a graph, where nodes represent tests and program entities, and edges capture coverage and code structure. A Gated Graph Neural Network learns features from this representation, followed by learning-to-rank techniques~\cite{lai18,chen09} to prioritize suspicious entities. Similarly, DeepRL4FL~\cite{DBLP:conf/icse/Li0N21a} frames FL as an image pattern recognition problem, combining vectorized representations of code coverage matrices, data dependencies, and source code with reinforcement learning and convolutional networks to discriminate between faulty and non-faulty entities.

While ML-based methods advance FL beyond SBFL, they often struggle to capture deep semantic relationships between code and natural language artifacts and to generalize across projects. To address these challenges, cross-project and transfer learning techniques have emerged. Zhang et al. introduced COOBA~\cite{CooBa}, an adversarial transfer learning approach for cross-project FL. COOBA extracts indicative public features across projects using a shared bug report encoder while preserving project-specific code features via individual extractors. Adversarial learning ensures effective extraction of shared information while minimizing negative transfer.

The recent emergence of transformer-based architectures and LLMs has further expanded FL capabilities. Transformers excel at capturing semantic relationships in code and natural language. Liang et al. proposed FLIM~\cite{liang2022modeling}, which splits source files into functions and uses a fine-tuned CodeBERT model~\cite{codebert2020} to compute similarity between bug reports and code functions. Function-level similarities are aggregated and combined with Information Retrieval (IR) features (e.g., bug-fixing frequency), and the resulting features are input to a learning-to-rank model to identify likely buggy files. Yang et al. proposed LLMAO\cite{DBLP:conf/icse/YangGMH24}, which fine-tunes a small set of bidirectional adapter layers on top of pre-trained LLMs to adapt them for line-level FL. Qin et al. introduced AGENTFL~\cite{DBLP:journals/corr/abs-2403-16362}, decomposing debugging into three steps (i.e., fault comprehension, codebase navigation, and fault confirmation) using multiple LLM-driven agents enhanced with prompt engineering to analyze failed test cases and validate suspicious methods. Li et al. proposed KEPT~\cite{kept25}, which fine-tunes the UniXcoder pre-trained LLM~\cite{guo2022unixcoder} by incorporating knowledge graphs. KEPT constructs knowledge graphs from historical documents and source code, then extracts relevant information from bug reports and change sets by aligning them with the graphs. This information is integrated into the LLM via soft-position embeddings~\cite{kbert2020} to preserve sentence structure and via a visible matrix to control token visibility, thereby generally enhancing contextual and domain-specific reasoning.

Despite these advances, most transformer and LLM-based approaches focus on FL in the SUT. Saboor et al.~\cite{saboor2025black} proposed the first LLM-based approach for test code fault localization, referred to as \baseline. This approach leverages execution logs to estimate test traces, which are combined with error messages to prompt the LLM and rank potential faulty locations. Baseline TCFL serves as our baseline because it is the only existing work targeting FL in test code and is described in detail in Section~\ref{sec:background}. 
Although \baseline improves effectiveness and scalability by pruning test code that is unlikely to be executed and hence irrelevant to the fault, it relies on general-purpose LLMs and does not incorporate external contextual knowledge. In contrast, \ourtool explicitly leverages accumulated debugging knowledge from CI pipelines. It utilizes previously observed fault-labeled test cases and their associated error messages to retrieve the most similar observed cases for a given query test. The identified faulty patterns are then used to annotate the query test code, guiding the LLM toward relevant regions and enabling more effective and context-aware FL.

Enhancing the contextual information provided to LLMs has become increasingly important across many domains, particularly due to the high cost and limited scalability of domain-specific fine-tuning, which often remains restricted to individual projects. In this context, RAG has emerged as an effective paradigm for improving LLM performance by incorporating external knowledge while maintaining efficiency ~\cite{rapgen23,Rag-tse25,code_completion}. 

In the context of FL, Shi et al proposed FaR-Loc~\cite{shi2025enhancing}, an LLM-based approach that integrates retrieval-augmented context to identify faulty function in the SUT. Given a failed test case and its associated runtime error traces, FaR-Loc first uses an LLM to generate a concise natural-language description of the failure behavior. This description is then used by a semantic dense retrieval component, which leverages a pre-trained code-understanding encoder to embed both the failure description and candidate methods into a shared semantic space. This enables the retrieval of methods with similar functional behavior that are likely related to the root cause. Finally, another LLM re-ranks the retrieved methods based on their contextual relevance. While this work demonstrates the benefits of incorporating retrieved contextual information for FL, it focuses on the SUT and assumes access to the full codebase, including test cases and coverage information. 

Our work distinguishes itself by (i) focusing on FL in system test code, (ii) leveraging accumulated debugging knowledge as the basis for a retrieval-augmented pipeline tailored to this setting, (iii) proposing a complete end-to-end framework for constructing, retrieving, and integrating such contextual knowledge into the LLM, and (iv) using a lightweight annotation mechanism to integrate this knowledge without adding computational overhead. We further conduct a comprehensive analysis of how different pipeline components affect both the effectiveness and efficiency of FL.

\section{Conclusions}\label{sec:conclusion}

In this paper, we propose \ourtool, a retrieval-based similar pattern annotation approach for LLM-based Test Code Fault Localization (TCFL). The core idea is to leverage a debugging knowledge memory accumulated through CI pipelines as an additional contextual guide for LLMs. This knowledge captures previously observed faulty patterns, which are used to enhance the identification of faulty elements in test code.

\ourtool introduces a pipeline that operationalizes this idea, incorporating multiple components and exploring the impact of various design choices on TCFL's efficiency and effectiveness across three real-world industrial datasets. Specifically, \ourtool first constructs a debugging knowledge memory from observed faulty test cases. Given a new faulty test query, it retrieves the most similar past cases. However, instead of directly injecting these cases as additional context into the LLM, \ourtool extracts similar faulty patterns (e.g., suspicious lines) and annotates them within the query test code. Designed for black-box environments typical of CI pipelines, \ourtool effectively reduces the fault localization search space while preserving the reasoning capabilities of off-the-shelf LLMs.

Comprehensive empirical evaluations across three industrial Python datasets demonstrate that \ourtool consistently improves FL effectiveness over existing LLM-based TCFL baselines while maintaining comparable efficiency. In particular, it improves top-1 Precision and Hit by 10.1--19.1 percentage points (pp), and Recall by 6.8--11.9 pp. These gains are consistent across different experimental settings and datasets. Additional analyses confirm the robustness and scalability of the approach under varying prompt designs, annotation thresholds, and filtering policies. Overall, this work demonstrates how additional contextual knowledge can be effectively defined, retrieved, and incorporated to improve the performance of TCFL using LLMs, enhancing effectiveness while preserving efficiency.
It also opens several directions for future research, including extending retrieval to multiple similar test cases, exploring alternative similarity measures and knowledge base construction strategies, investigating different annotation granularities, and incorporating richer contextual modalities such as code diffs or error patterns. Finally, evaluating the approach across additional programming languages and broader datasets would help assess its generalizability in diverse real-world scenarios.

\section*{Data and Tool Availability}\label{sec:availability}
Our tool implementation and data embeddings are available at \underline{
Available after review completion.}.
The source code of test cases and the contents of the log files are provided by our industry partner under a non-disclosure agreement and cannot be publicly released due to confidentiality constraints.

\section*{Acknowledgment}
This work was supported by a research grant from Huawei Technologies Canada Co., Ltd., as well as the Canada Research Chair and Discovery Grant programs of the Natural Sciences and Engineering Research Council of Canada (NSERC). Lionel C. Briand's contribution was partially funded by the Research Ireland grant 13/RC/209.

\bibliographystyle{ACM-Reference-Format}
\bibliography{references}

\end{document}